\documentclass[12pt]{iopjournal}

\usepackage{amssymb}
\usepackage{amsmath}
\usepackage{cancel}
\usepackage{xcolor}
\newcommand{\ext}[1]{\mathbf{d}\ensuremath{\mathbf{#1}}}
\newtheorem{thm}{Theorem}[section]
\newcommand{\rev}[1]{\textcolor{red}{#1}} 
\usepackage[normalem]{ulem}

\usepackage[pscoord]{eso-pic} 
\newcommand{\reportnumber}[1]{%
  \AddToShipoutPictureBG*{
    \AtPageUpperLeft{%
      \put(\dimexpr\paperwidth-2cm\relax,-4.7cm){
        \makebox[0pt][r]{\small\ttfamily Report No: #1}%
      }%
    }%
  }%
}

\begin{document}

\articletype{Paper} 

\reportnumber{PI/UAN-2026-740FT}

\title{Self-interacting $p$-forms and the shear-free anisotropic cosmological model}

\author{Nicolás Hernández$^{1,2}$\orcid{0009-0004-7974-7250}, Ben D. Normann$^3$\orcid{0000-0002-9508-7350}, and Yeinzon Rodríguez$^{1,4,*}$\orcid{0000-0002-9830-7128}}

\affil{$^1$Escuela de Física, Universidad Industrial de Santander, Ciudad Universitaria, Bucaramanga 680002, Colombia}

\affil{$^2$Departamento de Óptica, Centro de Investigación Científica y de Educación Superior de Ensenada, Carretera Ensenada-Tijuana No. 3918, Zona Playitas, Ensenada 22860, México}

\affil{$^3$Department of ICT and Natural Sciences, Norwegian University of Science and Technology, Postboks 1517 NO-6025, 6025 \r{A}lesund, Norway}

\affil{$^4$Centro de Investigaciones en Ciencias Básicas y Aplicadas, Universidad Antonio Nariño, Cra 3 Este \# 47A - 15, Bogotá D.C. 110231, Colombia}

\affil{$^*$Author to whom any correspondence should be addressed.}

\email{nhernandez@cicese.edu.mx, ben.d.normann@ntnu.no, yeinzon@saber.uis.edu.co}

\keywords{differential $p$-forms, anisotropic spacetimes, Bianchi and Kantowski-Sachs metrics, gauge-field theories.}

\begin{abstract}
 The cosmological principle may require more than observations demand: what is needed is isotropic expansion without vorticity. Extended Friedmann-Lemaitre-Robertson-Walker (eFLRW) cosmologies realise this with a non-vanishing anisotropic spatial curvature that is balanced by an anisotropic stress while the shear vanishes. Thorsrud and others investigated such scenarios for free differential $p$-forms.
 Here we extend the analysis to non-Abelian $p$-forms ($p=0,\dots,4$) with self interactions. The balance requires the anisotropic stress to decay as $a^{-2}$, $a$ being the cosmological scale factor, which we combine with the gauge-field continuity equation. The general results indicate that, in the presence of a potential, fine-tuning of the same is needed in order to balance the curvature. Notably, this is not the case for a $3$-form field $\underline{H}^a$ with a potential term $V(\underline{H}^2)=A+B\,\underline{H}^2$, where $A$ and $B$ are constants: in Bianchi class A, an uncharged 3-form balances the curvature with this potential without any tuning, and a charged one does so for a suitably chosen connection. This is the only candidate we have found where the potential does not require tuning. It is also an interesting candidate as it can be interpreted as a cosmological constant in combination with a dark-matter sector mimicking curvature.

\end{abstract}

\section{Introduction}

The cosmological principle is a cornerstone of mainstream cosmology;  it states that the universe is statistically homogeneous and isotropic at sufficiently large scales ($\gtrsim 100$ Mpc) \cite{Peter:2013avv,Weinberg:2008zzc,Mukhanov:2005sc}.  Having this principle in mind, modelling the universe in the framework of a pseudo-Riemannian manifold leads us to describe the spacetime at the background level by means of the FLRW metric \cite{Schutz:1980,MacCallum:1973kva}, which is subject to perturbations to describe the real properties and dynamics of the universe \cite{Lyth:2009imm}.

 The FLRW metric describes a universe in which expansion, but neither shear nor vorticity, is experienced by the congruence of fundamental observers \cite{Ellis:2012}.  If general relativity is taken to be the theory of gravity, the energy content of the universe, parametrized by the metric energy-momentum tensor, must then be described by a perfect fluid;  as a consequence, the energy content has vanishing momentum density and vanishing anisotropic stress. However, what astronomical observations really tell us is that the data are \textit{consistent with an isotropic and irrotational expansion}~\cite{Saadeh:2016sak}, which is equivalent to the absence of shear and vorticity.  Thus, a less stringent requirement on the energy content of the universe is that anisotropic stress may be present in the matter-energy content, but balanced by the anisotropic spatial curvature; this, in turn, allows the shear to vanish~\cite{Mimoso:1993ym}. This weaker version of the cosmological principle can then be implemented, at the zeroth-order level, by one of the homogeneous but anisotropic spacetimes, i.e., by one of the Bianchi or Kantowski-Sachs metrics \cite{MacCallum:1973kva,Ellis:2012,Zee:2016fuk}.

 Several works have thus far shown the existence of shear- and vorticity-free cosmologies with anisotropic stress and anisotropic spatial curvature that balance each other so that they exhibit background dynamics equivalent to that of the standard FLRW cosmology. Most notable is perhaps the investigation regarding shear-freeness led by Thorsrud \cite{Thorsrud:2019csi,Normann:2019auz,Thorsrud:2019clz,Normann:2017aav,Thorsrud:2017rpl}, who started from the $p$-form action
\begin{align}
\label{Thorsrudaction}
    S=-\frac{1}{2}\int{\mathcal{F}\wedge \star \mathcal{F}} \,,
\end{align}
where $\mathcal{F}$ is a differential $(p+1)$-form and $\star$ denotes the Hodge-star operator. The action in Eq.~(\ref{Thorsrudaction}) gives rise to source-free ``Maxwell-like'' equations of the form
\begin{align}
&\ext{}\star\mathcal{F}=0 \,, \quad\quad\textrm{Field equations} \,, \label{feqfree} \\
&\label{BianchiID}\ext{}\mathcal{F}=0 \,, \quad\quad\quad \textrm{Bianchi identity} \,, 
\end{align}
where $\ext{}$ is the covariant exterior space-time derivative, i.e., it includes the associated connection (Ricci rotation) coefficients. The first equation corresponds to the field equations and the second equation is the Bianchi identity, in this case following from the nilpotency of the exterior derivative operator:
$\mathcal{F} \equiv \ext{}\mathcal{K}$, where $\mathcal{K}$ is an underlying differential $p$-form.

 The $p$-form actions in the matter sector are well worth studying since they naturally give rise to gauge theories, allow for anisotropic stress, and can even account for the dark-matter phenomenon (see Refs. \cite{Plantier:2025hcm,Plantier:2026cgj}).  All these are desirable ingredients in a realistic model of the universe.

 In Ref. \cite{Thorsrud:2017rpl}, Thorsrud claimed that the only case capable of a dynamical cancellation of anisotropies is the case $p=0$ \footnote{Note that here and throughout we shall reference the $p$-form on the field level, and not on the corresponding field-strength level (the latter being a differential $(p+1)$-form).} (see also Ref. \cite{Carneiro:2001fz}), which, by Hodge duality, should be equivalent to the case $p=2$ (see also Ref. \cite{Koivisto:2010dr}) in the source-free case.\footnote{Except for an overall sign in the energy-momentum tensor.}  Interestingly, this case was shown to be stable under anisotropic perturbations \cite{Thorsrud:2019csi}.

 It is noteworthy that models claimed to be dynamically equivalent to FLRW cosmologies exist at all \cite{Thorsrud:2019csi}, and we are led to ask: are such extended FLRW (eFLRW) models realistic, or do they merely constitute a set of finely tuned special cases? This paper contributes to answering that question by extending the work of Thorsrud to what might be seen as a more realistic scenario where gauge theories and self-interactions are included; i.e., we consider interactions stemming both from a source term in the action and from the possible non-Abelian nature of the gauge fields involved.

  To summarize our results, Table~\ref{tab:summary} presents, for each $p$-form case, the necessary conditions for the balancing to work, together with the degree of fine tuning required. It is worthwhile noticing that Eqs.~(\ref{eq:FieldEq}), (\ref{eq:IntCond}) and (\ref{eq:J}) (to be introduced in the following section) will, in general, impose further restrictions and, hence, the final verdict in each positive case cannot be given yet.  Nevertheless, a particularly interesting non-trivial case that works {\it without any fine tuning} is the 3-form case where the potential involves a constant term as well as a quadratic term in the 3-form; this holds in Bianchi class A for a 3-form that is not charged under any non-Abelian Lie group, while a charged 3-form requires, in addition, a suitable choice of the connection $\mathcal{B}$. The role of interactions is of significant interest, as we regard the cosmological actions as approximations to the underlying particle-physics dynamics, where interactions are a crucial feature. Moreover, we agree with Thorsrud's findings in the appropriate limit.

\begin{table}[h]
\caption{\label{tab:summary} Summary of the shear-free balancing mechanism for each $p$-form case as dictated by the continuity equation. The fine-tuning entry is left blank where no balancing is possible.}
\centering
\footnotesize\setlength{\tabcolsep}{3pt}
\begin{tabular}{@{}lllcp{7.5cm}@{}}
\hline\hline
$p$ & Potential & Charge $\varrho$ & Is $\pi^\mathcal{K}=^3S$ non trivial? & Fine-tuning of $V(\mathcal{K}^2)$ or $\underline{\omega}^a$? \\
\hline
0 & $V=0$ & $\varrho=0$ & Yes & None. \\
0 & $V=0$ & $\varrho\neq0$ & Yes & $\underline{\omega}^a$ tuned by Eq.~(\ref{v3+1}). \\
0 & $V\neq0$ & $\varrho=0$ & Yes & $V$ tuned by Eq.~(\ref{nd-Vresult}) (or $V$ constant, $k=0$, and $\varrho \neq 0$); $\underline{\omega}^a$ tuned by Eq.~(\ref{nd-pfamily}). \\
0 & $V\neq0$ & $\varrho\neq0$ & Yes & $V$ tuned by Eq.~(\ref{nd-Vresult}) (or $V$ constant, $k=0$); $\underline{\omega}^a$ tuned by Eq.~(\ref{nd-pfamily}). \\
1 & $V=0$ & Any & No & \\
1 & $V\neq0$ & Any & Yes & $V$ tuned by Eq.~(\ref{cont1+1}). \\
2 & $V=0$ & $\varrho = 0$ & Yes & None. \\
2 & $V=0$ & $\varrho \neq 0$ & Yes & $\underline{\omega}^a$ tuned by Eqs. (\ref{phi3+1}) and (\ref{b3+1}). \\
2 & $V\neq0$ & Any & Yes & $V$ tuned by Eq.~(\ref{cont2+1}). \\
3 & $V=0$ & Any & No & \\
3 & $V\neq0$ & $\varrho = 0$ & Yes & None for $V$ first-degree in $\underline{H}^2$ (Bianchi class A); otherwise $V$ tuned by Eq.~(\ref{cont3+1a}). \\
3 & $V\neq0$ & $\varrho \neq 0$ & Yes & None for $V$ first-degree in $\underline{H}^2$ (Bianchi class A), with $\underline{\omega}^a$ chosen such that $f^a{}_{bc}\,\underline{\omega}^b(b^c)=0$; otherwise $V$ tuned by Eq.~(\ref{cont3+1a}). \\
4 & $V \neq 0$ & Any & No & \\
\hline\hline
\end{tabular}
\end{table}

 The rest of the paper is structured as follows. In Section~\ref{Sec:Nonabelian}, we explain our theoretical framework. In Section~\ref{ch:shear_condition}, we give a concise account of how to construct shear-free cosmologies 
by balancing the anisotropic spatial curvature with the anisotropic stress.  In Section~\ref{sec:non-Abelian}, which constitutes the main part of this work, we investigate the construction of shear-free cosmologies with underlying differential $p$-forms charged, in general, under some Lie group, and with a source term introduced in the action.  In Section~\ref{sec:conclusion}, we give our conclusions.

 Throughout the text, unless otherwise specified, we work in a non-coordinate orthonormal frame, Greek indices run from 0 to 3 and denote space-time components, and Latin indices run from $1$ to $\dim(\mathfrak{g})$, where $\mathfrak{g}$ is the Lie algebra associated with the gauge group, and denote gauge-field components.
We have employed the mostly plus signature for the space-time metric (see, e.g., Ref. \cite{Misner:1973prb}).

\section{Introducing interactions}
\label{Sec:Nonabelian}

 To explore the possibility of anisotropic hairs, it is essential to identify a matter source that can sustain anisotropies. A natural candidate for this role is the differential $p$-form, which provides a general framework for modelling anisotropic matter sectors. In particular, the construction of shear-free Bianchi or Kantowski-Sachs cosmologies needs a source capable of balancing the anisotropic spatial curvature that arises in the shear propagation equation, as discussed by Thorsrud~\cite{Thorsrud:2017rpl}.

 For a Lorentzian space-time manifold of dimension $n$, the canonical volume form $\eta$ is given by the relation $\eta=\star 1$, and hence any Lorentz scalar (function) $f$ defines a volume form $\mathcal{V} \equiv \star f$. Since the volume form is a top form, integrating it will again give a scalar. Therefore, a functional $S$ may be constructed in a coordinate-invariant manner as $S=\int \star f$. In order to extend Thorsrud's ideas to account for the character of gauge theories as well as self interactions, we consider a vector bundle where the base manifold is the 4-dimensional spacetime and the typical fibre is the Lie algebra $\mathfrak{g}$ associated to a specific Lie group.  We take $\mathcal{V}=-\frac{1}{2}\, \mathcal{F}^a\wedge \star \mathcal{F}_a - \star V(\mathcal{K}^2)$
where $\mathcal{F}^a$ are the components of a (Lie-algebra valued) field-strength differential $(p+1)$-form
\begin{equation}
    \mathcal{F}=\frac{1}{(p+1)!}\mathcal{F}^a_{\mu_1...\mu_{p+1}}\boldsymbol{t}_a\otimes\boldsymbol{\omega}^{\mu_1}\wedge...\wedge\boldsymbol{\omega}^{\mu_{p+1}} \,;
\end{equation}
in the above, $\boldsymbol{t}_a$ are the basis vectors of the fibre and $\boldsymbol{\omega}^{\mu}$ are the basis elements of the cotangent space. Meanwhile,
$\mathcal{K}^a$ are the components of a (Lie-algebra valued) underlying $p$-form gauge field
\begin{equation}
    \mathcal{K}=\frac{1}{p!}\mathcal{K}^a_{\mu_1...\mu_p}\boldsymbol{t}_a\otimes\boldsymbol{\omega}^{\mu_1}\wedge...\wedge\boldsymbol{\omega}^{\mu_p} \,,
\end{equation}
and $V(\mathcal{K}^2)$ is a scalar function of $\mathcal{K}^2$, defined below.

 We will construct $\mathcal{F}^a$ so that it is a differential form constructed from $\mathcal{K}^a$:
\begin{equation}
\label{PdK}
\mathcal{F}^a \equiv D\mathcal{K}^a \,,
\end{equation}
where $D$ is the gauge-covariant exterior derivative. For all adjoint-valued forms $\alpha$ other than the connection $\mathcal{B}$ itself, it is given by
\begin{equation}
\label{eq:DX}
D\alpha^a \equiv \ext{}\alpha^a + \varrho f^{a}{}_{bc}\,\mathcal B^b\wedge \alpha^c \,,
\end{equation}
where $\ext{}$ is the covariant exterior space-time derivative, $\varrho$ is the Lie group coupling constant, $f^a{}_{bc}$ are the structure coefficients of the Lie algebra, and 
\begin{equation}
\mathcal{B} = \mathcal{B}^a_\mu \boldsymbol{t}_a \otimes \boldsymbol{\omega}^\mu \,, 
\end{equation}
is the already mentioned 1-form connection on the fibre. Throughout this work, $\mathcal{B}$ is treated as an external, non-dynamical background field.

Let $u$ represent the unique time-like vector field describing the 4-velocity of a comoving observer with the fluid that represents the energy content of the universe. We may use $u$ to decompose $\mathcal{B}^a$ as\footnote{For further details regarding this 3+1 decomposition, refer to Appendix~\ref{App:decomposition}.}
\begin{align}
    \mathcal{B}^a=-\chi^a\, \underline{u}+\underline{\omega}^a \,, \quad \text{with} \quad \underline{\omega}^a(u) = 0 \,. \label{B3+1}
\end{align}
Gauge-freedom can always be used to set the projection of $\mathcal{B}$ along any chosen time-like congruence $u$ to zero (``temporal gauge'') without any loss of generality; we prove this fact in Appendix~\ref{App:TemporalGauge}:
\begin{align}
    \label{chi0}
    \chi^a=0 \,, \quad\textrm{Temporal gauge.}
\end{align}
Nevertheless, we choose to keep $\chi^a$ in the equations, rather invoking the temporal gauge explicitly whenever needed. On the other hand, the curvature associated to the connection $\mathcal{B}$ is given by
\begin{equation}
\label{eq:curvature}
D\mathcal{B}^a \equiv \ext{}\mathcal{B}^a+\frac{\varrho}{2} f^{a}{}_{bc}\,\mathcal{B}^b\wedge\mathcal{B}^c \,.
\end{equation} 
Since $\mathcal{B}$ is a genuine connection, it does not transform tensorially (in the adjoint representation) under a gauge transformation of the bundle.

Having said this, we start by considering the action
\begin{equation}
    S=\int \left[-\frac{1}{2}\mathcal{F}^a\wedge \star \mathcal{F}_a - \star V(\mathcal{K}^2) 
    \right] \,,
\end{equation}
where $\mathcal{K}^2 \equiv \mathcal{K}_a\mathcal{K}^a$, the Latin index being lowered by the Cartan-Killing metric, $G_{ab}$. Lorentz and gauge invariance require $V$ to depend on $\mathcal{K}^a$ only through invariants, i.e., full contractions of the space-time indices that are invariant under the adjoint action of the gauge group. The simplest such invariant, and the only one built from a single isovector when $\mathfrak{g}=\mathfrak{su}(2)$, is $\mathcal{K}^2$; we restrict ourselves to potentials of the form $V(\mathcal{K}^2)$ throughout. A prime on $V$ denotes a derivative with respect to its argument,
$V'\equiv \mathrm{d}V/\mathrm{d}(\mathcal{K}^2)$. We assume throughout that the gauge Lie algebra $\mathfrak{g}$ is semisimple and compact,\footnote{A semi-simple Lie algebra admits a positive-definite invariant metric if and only if it is compact~\cite{Knapp:2002}. For a non-compact $\mathfrak{g}$ the Cartan-Killing metric $G_{ab}$ is indefinite, so the gauge-invariant kinetic term $-\frac12\mathcal{F}^a\wedge\star\mathcal{F}_a$ has negative-norm directions in isospin space; in this case, the classical energy density is not positive and, upon quantisation, ghost modes (states of negative norm) rise and the bound $-1<w<1$ below is lost. We therefore regard a non-compact $\mathfrak{g}$ as unphysical for the matter sector and exclude it.} so that $G_{ab}$ is non degenerate and positive definite and the fully-lowered structure constants $f_{abc}\equiv G_{ad}f^{d}{}_{bc}$ are totally antisymmetric (a property used explicitly later in this section -- see the nice discussion in Chapter 15 of Ref. \cite{Weinberg:1996kr}).
This action gives rise to ``Yang-Mills-like'' equations of the form
\begin{align}
&D\star\mathcal{F}^a= (-1)^p \star \frac{\partial V}{\partial \overline{\mathcal{K}}_a}
\,, \quad\quad \textrm{Field equations} \,, \label{eq:FieldEq}\\
&D \mathcal{F}^a
=
\varrho f^{a}{}_{bc}\,
D\mathcal{B}^b \wedge \mathcal{K}^c  \,, \quad\quad \hspace{-1mm} \textrm{Integrability condition} \,. \label{eq:IntCond}
\end{align}
Here we have introduced $\overline{\mathcal{K}}^a$ as the $p$-vector dual to $\mathcal{K}^a$ by metric $g$-duality ($g$ being the space-time metric). The integrability condition above replaces the Bianchi identity, Eq.~(\ref{BianchiID}), in the presence of a non-vanishing curvature, both being derivable from the Ricci identity. Notice the failure of the nilpotency of the exterior derivative operator $D$ whenever the curvature of the typical fibre is non-vanishing: $D\mathcal{F}=D^2\mathcal{K}\neq 0.$ Apart from its appearance in the consistency condition in Eq.~(\ref{eq:J}) below, neither the curvature of the typical fibre nor the integrability condition are used in this work; they are presented here for the sake of completeness and for their connection to the aforementioned uncharged-under-any-Lie-group case studied by Thorsrud \cite{Thorsrud:2017rpl}.

 Here $\mathcal{B}$ is treated as an external, non-dynamical background: on shell, diffeomorphism invariance of the action then implies
\begin{equation}
\nabla_\mu \mathcal{T}^{\mathcal{K}\,\mu}{}_{\nu}=\varrho\; j_b^{\mu}\,(D\mathcal{B}^b)_{\mu\nu}\,,\qquad j_b\equiv\star J_b\,,\quad J_b\equiv f_{bca}\,\mathcal{K}^c\wedge\star\mathcal{F}^a\,, \label{eq:ConsB}
\end{equation}
where $\mathcal{T}^\mathcal{K}$ is the energy-momentum tensor associated to the $p$-form, $j_b$ is the isospin current 1-form, and $j_b^{\mu}$ are its components with the index raised by $g$ (so $J_b=\star j_b$). Since Einstein's equations, through the contracted Bianchi identity $\nabla^\mu G_{\mu\nu}=0$ (where $G_{\mu \nu}$ are the components of the Einstein tensor), require $\nabla_\mu \mathcal{T}^{\mu}{}_{\nu}=0$, with $\mathcal{T}$ being the total energy-momentum tensor (see Section~\ref{ch:shear_condition}), and since the remaining matter content is assumed to be separately conserved, every solution must in addition satisfy
\begin{equation}
\varrho\; j_b^{\mu}\,(D\mathcal{B}^b)_{\mu\nu}=0\,. \label{eq:J}
\end{equation}
In the uncharged case ($\varrho=0$) or for a flat connection ($D\mathcal{B}=0$), the previous expression is satisfied automatically; note, however, that a flat connection is locally pure gauge, so that this case is physically equivalent to the uncharged one. If $\mathcal{B}$ were dynamical, total conservation would instead be guaranteed by its own field equations: without a kinetic term, these impose $J_b=0$, while with a Yang--Mills term the exchange is compensated by $\mathcal{B}$'s own energy-momentum tensor. In our case, however, $\mathcal{B}$ is an external, fixed background, and, therefore, Eq.~(\ref{eq:J}) is required. Like the field equations~(\ref{eq:FieldEq}) and the integrability condition~(\ref{eq:IntCond}), this condition is not imposed on the configurations studied in Section~\ref{sec:non-Abelian}; it is trivially satisfied whenever $\varrho=0$ or $\mathcal{B}=0$, and is left for future work (see Section~\ref{sec:conclusion}).

\section{Balancing the anisotropic spatial curvature with the anisotropic stress tensor}
\label{ch:shear_condition}

 The Bianchi cosmological models are characterised by metrics supporting a three-dimensional isometry group, which acts simply transitively on space-like hypersurfaces, known as surfaces of homogeneity \cite{MacCallum:1973kva,Ellis:2012}.\footnote{The Kantowski-Sachs spacetime belongs to a class of locally rotationally symmetric spacetimes with a three-dimensional subgroup of isometries that acts multiply transitively on two-dimensional hypersurfaces of maximal symmetry \cite{MacCallum:1973kva}.  This is the reason why the Kantowski-Sachs metric falls outside the Bianchi classification.  Nevertheless, the conditions for a shear-free cosmology using the Kantowski-Sachs metric are the same as the ones employing a Bianchi metric as proved in Ref. \cite{Thorsrud:2017rpl}.} Under this structure and the application of Frobenius' theorem (see Ref. \cite{lee2003introduction}), these spacetimes allow for a foliation into homogeneous surfaces, denoted as $\Sigma_t$, each identified by the cosmic time coordinate $t = x^0$. \\

 We shall focus on the non-tilted models, i.e., those where $u$ is orthogonal to the foliation of homogeneous surfaces.    
In these models, the congruence of fundamental observers is characterised by being non-accelerated and irrotational, with the only kinematic quantities of interest being the Hubble expansion scalar $H$ and the shear tensor $\sigma$, given by 
\begin{align}
    3 H \equiv \nabla \cdot u \,, \quad\quad \text{and} \quad\quad \sigma \equiv {\rm Sym}(\nabla \underline{u}) -H h \,,
\end{align}
respectively.  Notice that Sym means normalized symmetrization, $\underline{u}$ is the 1-form dual to the four-velocity $u$ by metric $g$-duality, and $h \equiv g + \underline{u} \otimes \underline{u}$ \cite{Gourgoulhon:2012ffd}. \\

 Besides, we shall consider two categories of matter fields: the constituents of the $\Lambda$ Cold Dark Matter ($\Lambda$CDM) model, represented as a collection of comoving perfect fluids, and a \( p \)-form gauge field, denoted as \( \mathcal{K} \). To avoid inducing anisotropy in the time direction, we impose the requirement that the heat flux generated by the gauge field vanishes:
\begin{equation}
    q =0 \,.
\end{equation}
Consequently, the energy-momentum tensors for the whole $\Lambda$CDM fluid, \( {\cal T} \), and for the gauge field, \( {\cal T}^{{\cal K}} \), assume the following form:
\begin{align}
    &{\cal T} =\rho \ \underline{u} \otimes \underline{u} +P h \,, \quad\quad\quad\quad \hspace{1mm} \text{where} \quad\quad \rho=\sum_{\ell} \rho_{\ell} \,, \quad \text{and} \quad P=\sum_{\ell} P_{\ell} \,,\\
    &{\cal T}^{\mathcal{K}}=\rho^{\mathcal{K}} \ \underline{u} \otimes \underline{u} +P^{\mathcal{K}} h +\pi^\mathcal{K} \,, 
\end{align}
where the index $\ell$ runs over all $\Lambda$CDM constituent fluids, $\rho^{(\mathcal{K})}$ is the energy density of the whole $\Lambda$CDM (gauge-field) fluid, $P^{(\mathcal{K})}$ is the isotropic pressure of the whole $\Lambda$CDM (gauge-field) fluid, and $\pi^{\mathcal{K}}$ is the anisotropic stress of the gauge-field fluid. Thus, the total energy-momentum tensor takes the form
\begin{align}
    T=\left(\rho^{\mathcal{K}}+\rho\right) \underline{u} \otimes  \underline{u}+\left(P^{\mathcal{K}}+P\right) h +\pi^\mathcal{K} \,.
\end{align}  \\

 The gravitational field equations dictate that the matter and gauge fields satisfy the evolution equations
\begin{align}
    &\dot{\rho}+3 H(\rho+P)=0 \,, \\
    &\dot{\rho}^{\mathcal{K}}+3 H\left(\rho^{\mathcal{K}}+P^{\mathcal{K}}\right)=- {\rm Cont} (\pi^\mathcal{K} \otimes \sigma)\label{eq:4.6} \,,
\end{align}
where an overdot denotes a time derivative along the congruence -- such as $\dot{\rho}=u \cdot \nabla \rho$ -- and Cont means contraction. 
Moreover, the congruence evolves according to the Raychaudhuri equation and the shear propagation
equation in addition to the Friedmann equation that constrains the variables: 
\begin{align}
    &\dot{H}+H^2=-\frac{1}{6}(\rho+3 P)-\frac{1}{6}\left(\rho^{\mathcal{K}}+3 P^{\mathcal{K}}\right)-\frac{2}{3} \sigma^2\label{eq:4.7} \,,\\
    &\dot{\sigma} +3 H \sigma =\pi^\mathcal{K}-{ }^3 S \,, \label{eq:4.8} \\
    &3 H^2-\sigma^2+\frac{{ }^3 R}{2}=\rho+\rho^{\mathcal{K}}\label{eq:4.9} \,.
\end{align}
Here, the shear scalar $\sqrt{|\sigma^2|}$ is defined by $\sigma^{2}= {\rm Cont} (\sigma \otimes \sigma)$, and 
\begin{align}
    { }^3 S \equiv { }^3 {\rm Ricci}-\frac{{ }^3 R}{3} h \,,
\end{align}
represents the trace-free part of the three-dimensional Ricci tensor $^{3}$Ricci defined on $\Sigma_{t}$ (of course, $^3R$ is the three-dimensional Ricci scalar). It is clear from the previous expression that $^3 S$ is the anisotropic curvature of the homogeneous slices. \\

 In general, Eqs. (\ref{eq:4.7})-(\ref{eq:4.9}) have more unknowns than equations, unless the required equations of state are specified. Therefore, we require a matter-gauge field model with suitable equations of state that can sustain the inherent anisotropy in Bianchi or Kantowski-Sachs models.\\

 Having in mind the considerations in the introduction, we wish a cosmological model with vanishing shear tensor during the whole evolution. Consequently, the vanishing of the shear and of its time derivative reduces the shear propagation expression in Eq. (\ref{eq:4.8}) to
\begin{align}\label{eq:4.11}
    { }^3 S= \pi^\mathcal{K} \,,
\end{align}
which is known as the shear-free condition \cite{Mimoso:1993ym}.\\

  The main goal of this paper consists in systematically investigating under what conditions the introduction of self interactions in the action describing the underlying $p$-form gauge field $\mathcal{K}$ is capable of balancing the anisotropic spatial curvature in this way, and what role the possible non-Abelian character of the fields plays, in contrast to the free-single-field case investigated in Ref. \cite{Thorsrud:2017rpl}.  \\

 Before moving to the next section, we will enunciate the following theorem \cite{Thorsrud:2017rpl} that will be key to the subsequent developments:

\begin{thm}{\rm {\bf (Anisotropic spatial curvature decay in Bianchi models with conformal expansion)}} 
    The components of the anisotropic spatial curvature \( {}^3 S \) relative to an orthonormal basis exhibit a decay proportional to \( 1/a^2 \) in the shear-free limit, where \( a(t) \) denotes the scale factor governing the conformal expansion.  \label{1sttheo}
\end{thm}

\subsection*{{\bf Proof}}

 The proof of this theorem is given in the arXiv version 2 of Ref. \cite{Thorsrud:2017rpl}, and we reproduce it here in Appendix~\ref{App:Knownth}.\\

\section{Attempts towards balancing the anisotropic curvature with the anisotropic stress}
\label{sec:non-Abelian}

  We now proceed to examine the non-trivial $p$-form gauge-field configurations in spacetime for $p \in \{0,1,2,3,4\}$, when the $p$-form gauge fields are, in general, members of a Lie algebra, as introduced in Section \ref{Sec:Nonabelian}.

\subsection{$p=0$ case}
\label{sub:0-form_non}
 We start with the following action for an underlying differential 0-form $\phi^a$: 
\begin{equation}
    S=\int \left[-\frac{1}{2}\mathcal{F}^a\wedge \star \mathcal{F}_a- \star V(\phi^2) \right] \,.
\end{equation} \\
 The 1-form  ${\cal F}^a$ can be decomposed as\footnote{For further details regarding this 3+1 decomposition, refer to Appendix~\ref{App:decomposition}.}
\begin{align}
    \mathcal{F}^a=-\varphi^a\, \underline{u}+\underline{v}^a \,, \quad \text{with} \quad \underline{v}^a(u) = 0 \,.
\end{align}
Here, $\varphi^a$ is a scalar field that satisfies $\varphi^a = \mathcal{F}^a(u)$, $\underline{u}$ is the 1-form dual to $u$ by metric $g$-duality, and $\underline{v}^a = \mathcal{F}^a + \mathcal{F}^a(u) \ \underline{u}$. Hence, the energy-momentum tensor associated to this gauge field takes the form 
\begin{align}\label{eq:R-3}
    \mathcal{T}^\mathcal{K}=& \left[\frac{1}{2}\,(\varphi^{2}+v^{2}) + V(\phi^2) \right] \, \underline{u} \otimes \underline{u} \nonumber \\
    &-2\varphi^{a}\, {\rm Sym}(\underline{u} \otimes \underline{v}_a) \nonumber \\
    &+ \left[\frac{1}{2}\,(\varphi^{2}-v^{2})-V(\phi^2) \right]\,h \nonumber \\
    &+\underline{v}_a \otimes \underline{v}^a \,,
\end{align}
where $\varphi^{2} \equiv \varphi_a\,\varphi^a$ and $v^{2} \equiv \underline{v}^a (v_a)$, $v_a$ being the vector dual to $\underline{v}_a$ by metric $g$-duality. \\
 Given these conditions, the energy density, pressure, heat flux, and anisotropic stress associated with the energy-momentum tensor in Eq. (\ref{eq:R-3}) are given by
\begin{align}\label{eq:R-4}
    &\rho^\mathcal{K}=\frac{1}{2}\left(\varphi^2 + v^2\right) + V(\phi^2) \,, \nonumber \\ &P^\mathcal{K}=\frac{1}{2}\left(\varphi^2 -\frac{1}{3} v^2\right) - V(\phi^2) \,, \nonumber \\
    & q^\mathcal{K} =-\varphi^{a}\, \underline{v}_{a} \,, \nonumber \\ &\pi^\mathcal{K}=\underline{v}^a \otimes \underline{v}_a-\frac{1}{3} v^2 h \,.
\end{align}
We observe that, for $\rho>0$ and $V\geq0$, this matter sector's equation-of-state parameter $w$ is confined to the intervall $-1<w<1$ depending on the values of the fields and of the potential.\\

 It is worth stressing that 
\begin{equation}
\mathcal{F}^a \equiv \ext{}\phi^a + \varrho f^a{}_{bc} \mathcal{B}^b \wedge \phi^c \,,
\end{equation}
so that
\begin{align}
&\varphi^a = \ext{}\phi^a (u) + \varrho f^a{}_{bc} \chi^b \phi^c \,, \label{vphi3+1} \\
&\underline{v}^a = \ext{}\phi^a + \ext{}\phi^a(u) \underline{u} + \varrho f^a{}_{bc} \underline{\omega}^b \phi^c \,, \label{v3+1} 
\end{align}
where we have decomposed $\mathcal{B}^a$ as
\begin{align}
    \mathcal{B}^a=-\chi^a\, \underline{u}+\underline{\omega}^a \,, \quad \text{with} \quad \underline{\omega}^a(u) = 0 \,.
\end{align}
The time evolution of the different quantities involved in these expressions depends on the existence of the potential term in the action.

\subsubsection{$V = 0$} \label{0ffree}
 In this case, the homogeneity requirement comes in on the field-strength level; $\mathcal{F}^a$ must be homogeneous, and hence $\varphi^a$ and $\underline{v}^a$ must too. Notice, however, that the 0-form $\phi^a$ needs not be homogeneous.  Eqs. (\ref{vphi3+1}) and (\ref{v3+1}) are, therefore, preserved. \\
 On the other hand, to make $q^\mathcal{K} = 0$ without making $\pi^\mathcal{K} = 0$, we need $\varphi^a \perp \underline{v}_a$ in the isospin space with $\underline{v}_a \neq 0$.  In such a case, as described in Sec. \ref{ch:shear_condition}, the components of $\pi^\mathcal{K}$ in an orthonormal basis evolve as $a^{-2}$ in the shear-free limit (see Eq. (\ref{eq:4.11}) and Theorem \ref{1sttheo});  this, in turn, implies that $\underline{v}^a \propto a^{-1}$ according to the expression for $\pi^\mathcal{K}$ in Eq. (\ref{eq:R-4}). \\

Let us study the free case in the first place, i.e., let us assume $\varrho = 0$.  Then, Eqs. (\ref{vphi3+1}) and (\ref{v3+1}) reduce to
\begin{align}
&\varphi^a = \ext{}\phi^a (u) \,, \label{vphi3+1free} \\
&\underline{v}^a = \ext{}\phi^a + \ext{}\phi^a(u) \underline{u} \,. \label{v3+1free} 
\end{align}
According to Eq. (\ref{v3+1free}), $\underline{v}^a \propto a^{-1}$ as long as $\dot{\phi}^a = 0$.\footnote{The spatial derivative operator in the orthonormal-frame formalism carries a factor $1/a$ on its own, as described in Appendix~\ref{App:Tetrad}.}  However, this implies $\varphi^a = 0$ according to Eq. (\ref{vphi3+1free}). \\

Thus, the time evolution of the different terms that compose the gauge-field continuity equation is given by
\begin{align}
&\dot{\rho}^\mathcal{K} = -v_0^2 \frac{\dot{a}}{a^3}  \,, \nonumber \\
&3H(\rho^\mathcal{K} + P^\mathcal{K}) = \frac{\dot{a}}{a} v_0^2 \frac{1}{a^2} \,,
\end{align}
which means this equation is automatically satisfied.  In addition, $\varphi^a$ being zero, $q^\mathcal{K} = 0$ automatically.  No fine tuning is required and Thorsrud's result in Ref. \cite{Thorsrud:2017rpl} is reproduced. \\

Now, let us study the non-free case. According to Eq. (\ref{v3+1}), $\underline{v}^a \propto a^{-1}$ as long as $\dot{\phi}^a = 0$ and $\underline{\omega}^b \propto a^{-1}$.  This, in turn, and according to Eq. (\ref{vphi3+1}), implies that $\varphi^a$ evolves as $\chi^a$ does. \\

The time evolution of the different terms that compose the gauge-field continuity equation is given by
\begin{align}
&\dot{\rho}^\mathcal{K} = \varphi \dot{\varphi} -v_0^2 \frac{\dot{a}}{a^3}  \,, \nonumber \\
&3H(\rho^\mathcal{K} + P^\mathcal{K}) = 3 \frac{\dot{a}}{a} \left(\varphi^2  +\frac{1}{3} v_0^2 \frac{1}{a^2}\right) \,,
\end{align}
where we can see that the gauge-field continuity equation, in the shear-free limit, is satisfied as long as 
\begin{equation}
\varphi^a(\dot{\varphi}_a + 3H\varphi_a) = 0 \,. \label{notrouble}
\end{equation} \\

The latter condition is possible to satisfy;  in fact, since there is no a predetermined evolution for $\chi^a$, Eq. (\ref{notrouble}) does not impose
 a particular evolution for the scale factor that cannot be possible to satisfy for an arbitrary matter content in the $\Lambda$CDM fluid.  The condition $\varphi^a \perp \underline{v}_a$ is possible to satisfy by tuning the initial conditions in $\varphi^a$ and $\underline{v}_a$; such tuning is unnecessary if $\varphi^a$ is chosen to vanish although this itself would require some fine tuning according to Eq. (\ref{vphi3+1}) if $\phi^a$ is charged under some non-Abelian Lie group.  Anyway, some fine tuning is necessary to make $\underline{\omega}^a \propto a^{-1}$ which might be  unrealizable once the kinetic term for $\mathcal{B}^a$ and other interaction terms are added to the Lagrangian. \\
 The final conclusion is that the realization of the shear-free cosmology when the field in charge of producing the balance between the anisotropic curvature and the anisotropic stress is a 0-form that is charged under some non-Abelian Lie group and whose potential vanishes is always possible although some fine tuning is necessary.
 
\subsubsection{$V \neq 0$}
 The presence of a potential term in the Lagrangian together with the homogeneity of space imposes the constraint $\ext{}\phi^a + \ext{}\phi^a(u) \underline{u} = 0$ on the 0-form.  Then, the expression in Eq. (\ref{v3+1}) reduces to
\begin{equation}
\underline{v}^a = \varrho f^a{}_{bc} \underline{\omega}^b \phi^c \,. \label{nd-vcharged}
\end{equation} \\
 The $\pi^\mathcal{K}$ requirement still needs $\underline{v}^a \propto a^{-1}$ which, according to the previous expression, implies $\underline{\omega}^b \phi^c \propto a^{-1}$.  We are going to (reasonably) consider the one-parameter family of ansätze
\begin{equation}
    \phi^a=\phi_0^a\,a^k \,, \qquad \underline{\omega}^b=\omega_0^b\,a^{-1-k} \,, \label{nd-pfamily}
\end{equation}
for a real constant $k$ and a constant isovector $\phi_0^a$. For any $k$, Eq.~(\ref{nd-vcharged}) gives $\underline{v}^a\propto a^{-1}$ directly, as required. With $\chi^a$ kept, Eq.~(\ref{vphi3+1}) gives
\begin{equation}
    \varphi^a = k H \phi^a + \varrho f^a{}_{bc}\chi^b\phi^c = a^k\left(k H \phi_0^a + \varrho f^a{}_{bc}\chi^b\phi_0^c\right) \,. \label{nd-varphi-chi}
\end{equation}
Since $H$ has a definite time dependence, the natural way to give $\varphi^a$ one too is by requiring $\chi^a=\chi_0^aH$, with $\chi_0^a$ a constant isovector (its sign is absorbed in $\chi_0^a$). Then
\begin{equation}
    \varphi^a = H a^k \Lambda^a \,, \qquad \Lambda^a\equiv k\phi_0^a+\varrho f^a{}_{bc}\chi_0^b\phi_0^c \,, \label{nd-Lambda}
\end{equation}
with $\Lambda^a$ being a constant. There are two possibilities:
\begin{align}
&\varphi^a \propto a^{k-1}\dot a \quad (\Lambda^a\neq0) \,, \label{nd-opt1}\\
&\quad\quad\quad\quad\quad\;\lor \nonumber \\
&\varphi^a = 0 \quad (\Lambda^a=0) \,. \label{nd-opt2}
\end{align}
Contracting $\Lambda^a$ with $\phi_{0a}$ and using the total antisymmetry of $f_{abc}$ gives $\Lambda^a\phi_{0a}=k\,\phi_0^2$. The second possibility, $\Lambda^a=0$, therefore requires $k\,\phi_0^2=0$ and, since $G_{ab}$ is positive-definite, this means $k=0$, i.e.,\ $\phi^a$ is constant with $f^a{}_{bc}\chi_0^b\phi_0^c=0$.\\

 As discussed in Section~\ref{Sec:Nonabelian}, we take $V=V(\phi^2)$, with $\phi^2\equiv\phi^a\phi_a\propto a^{2k}$. Since $\underline{v}^a\propto a^{-1}$, the $v^2$ terms in the continuity equation cancel exactly as in the $V=0$ case above. Using $\dot\phi^a=kH\phi^a$, what remains is
\begin{equation}
    \varphi_a\dot\varphi^a+3H\varphi^2+2kH\,V'(\phi^2)\,\phi^2=0 \,, \label{nd-Veq}
\end{equation}
where the prime means a derivative with respect to $\phi^2$.
For $\Lambda^a\neq0$ and $k\neq0$, inserting Eq.~(\ref{nd-Lambda}) and simplifying gives
\begin{equation}
    V'(\phi^2)=-\frac{\Lambda^2}{2k\,\phi_0^2}\left[\dot{H}+(k+3)H^2\right] \,, \label{nd-Vresult}
\end{equation}
with $\Lambda^2\equiv\Lambda^a\Lambda_a$. For completeness we note that the temporal gauge, $\chi^a=0$, is contained as the special case $\chi_0^a=0$; thus $\Lambda^a=k\phi_0^a$ so that $\varphi^a=kH\phi^a$ (the isospin direction of $\phi^a$ is fixed) and Eq.~(\ref{nd-Vresult}) reduces to $V'(\phi^2)=-\frac{k}{2}\dot H-\frac{k(k+3)}{2}H^2$ which is further reduced to $V'(\phi^2)=H^2+\dot H/2$ for $k=-1$. More generally: given {\it any} desired background evolution $H(t)$ and {\it any} choice of $k\neq0$ and $\Lambda^a\neq0$, Eq.~(\ref{nd-Veq}) determines the potential $V(\phi^2)$ needed to support it. For $k=0$ and $\Lambda^a\neq0$, the potential drops out and Eq.~(\ref{nd-Veq}) requires $\dot H=-3H^2$ instead which cannot be possible to satisfy for an arbitrary matter content in the $\Lambda$CDM fluid. In the case $\Lambda^a=0$, one has $k=0$ so that $\phi^2$ is constant and $V$ is flat; Eq.~(\ref{nd-Veq}) is then satisfied with no further condition on the potential but $\phi$ is required to be charged under some non-Abelian Lie group, i.e., $\varrho \neq 0$, to prevent the kinetic term in the action from vanishing.\\

 Finally, we will check the requirement $q^\mathcal{K}=0$, i.e.\ $\varphi^a\underline{v}_a=0$. The term $kH\phi_0^a$ in $\Lambda^a$ drops out by the total antisymmetry of $f_{abc}$, leaving the constant condition
\begin{equation}
\varrho^2 f_{ade}\chi_0^d\phi_0^e\, f^a{}_{bc}\,\omega_0^b\phi_0^c=0 \,. \label{nd-q0}
\end{equation}
This condition is identically satisfied for $\chi_0^a=0$ but, otherwise, it is a condition on the constant isovectors $\chi_0^a$, $\phi_0^a$ and $\omega_0^a$ that can be met by their choice. \\

 The final conclusion is that the presence of a potential term for the 0-form is consistent with a shear-free cosmology although some fine tuning is required.


\subsection{$p=1$ case}

 We start with the following action for an underlying differential 1-form $\underline{A}^a$:
\begin{equation}
    S=\int \left[-\frac{1}{2}\mathcal{F}^a\wedge \star \mathcal{F}_a - \star V(A^2) \right] \,.
\end{equation} \\

 Considering the 3+1 decomposition\footnote{For further details regarding this decomposition, refer to Appendix~\ref{App:decomposition}.}, we arrive at 
\begin{align}
    &{\cal F}^a = \underline{u} \wedge \underline{p}^a +\star (\underline{u} \wedge \underline{r}^a) \,, \quad \text{with} \quad \underline{p}^a (u) =0 \quad\text{and}\quad \underline{r}^a (u) =0 \,, \label{F3+1} \\
    &\underline{A}^a = -\varphi^a\, \underline{u}+\underline{v}^a \,, \quad \text{with} \quad \underline{v}^a(u) = 0 \,. \label{A3+1}
\end{align}
In these expressions, $\underline{p}^a = \mathcal{F}^a (\cdot,u)$, $\underline{r}^a = (\star \mathcal{F}^a) (u,\cdot)$, $\varphi^a = \underline{A}^a(u)$, and $\underline{v}^a = \underline{A}^a + \underline{A}^a(u) \ \underline{u}$. \\

 Hence, the energy-momentum tensor associated to this gauge field takes the form 
\begin{align}\label{eq:R-9}
    \mathcal{T}^\mathcal{K} = &\underline{u} \otimes \underline{u} \left[\frac{1}{2}(p^2 + r^2) + 2V'(A^2) \varphi^2 + V(A^2) \right] \nonumber \\
    &+2 \ {\rm Sym}\{\underline{u} \otimes [\star(\underline{p}^a \wedge \underline{r}_a)](u,\cdot)\} - 4V'(A^2)\varphi^a {\rm Sym}(\underline{u} \otimes \underline{v}_a) \nonumber \\
    &+ h \left[\frac{1}{2}(p^2 + r^2) - V(A^2)\right] \nonumber \\
    &-\underline{p}^a \otimes \underline{p}_a -\underline{r}^a \otimes \underline{r}_a+2V'(A^2) \underline{v}^a \otimes \underline{v}_a \,.
\end{align}
In the above, $p^2 \equiv \underline{p}^a (p_a)$ and $r^2 \equiv \underline{r}^a(r_a)$, where $p_a$ and $r_a$ are the vectors dual to $\underline{p}_a$ and $\underline{r}_a$ by metric $g$-duality, respectively, and the prime means a derivative with respect to $A^2 \equiv \underline{A}^a (A_a)$, $A_a$ being the vector dual to $\underline{A}_a$ by metric $g$-duality. \\

 Given these conditions, the energy density, pressure, heat flux, and anisotropic stress
associated with the energy-momentum tensor in Eq. (\ref{eq:R-9}) take the form 
\begin{align}
    \rho^\mathcal{K} &=\frac{1}{2}(p^2 + r^2) + 2V'(A^2) \varphi^2 + V(A^2)  \,, \nonumber \\
    P^\mathcal{K} &=\frac{1}{6} (p^2+ r^2) - V(A^2) + \frac{2}{3} V'(A^2) v^2 \,, \nonumber \\
    q^\mathcal{K} &= [\star(\underline{p}^a \wedge \underline{r}_a)](u,\cdot) -2V'(A^2) \varphi^a \underline{v}_a  \,, \nonumber \\
    \pi^\mathcal{K} &=-\underline{p}^a \otimes \underline{p}_a -\underline{r}^a \otimes \underline{r}_a + 2V'(A^2) \underline{v}^a \otimes \underline{v}_a + h \left[\frac{1}{3}(p^2+r^2) - \frac{2}{3} V'(A^2) v^2\right]  \,. \label{matter2-f}
\end{align} \\

 Let us remember Eq. (\ref{PdK}), i.e.,
\begin{equation}
\mathcal{F}^a \equiv \ext{}\underline{A}^a + \varrho f^{a}{}_{bc}\,\mathcal B^b\wedge \underline{A}^c \,, \label{1gfst}
\end{equation}
where the connection on the fibre, $\mathcal{B}$, is a 1-form and, therefore, admits a $3+1$ decomposition as described in Appendix \ref{App:decomposition}:
\begin{equation}
\mathcal{B}^a = - \chi^a \underline{u} + \underline{\omega}^a \,, \quad \text{with} \quad \underline{\omega}^a (u) = 0 \,. \label{B3+1}
\end{equation}
Combining Eqs. (\ref{A3+1}), (\ref{1gfst}), and (\ref{B3+1}), we end up with
\begin{align}
\mathcal{F}^a = \underline{u} \wedge [\ext{} \varphi^a - \varrho f^a{}_{bc} (\chi^b \underline{v}^c - \varphi^c \underline{\omega}^b)] + \ext{}\underline{v}^a + \varrho f^a{}_{bc} \underline{\omega}^b \wedge \underline{v}^c \,.
\end{align}
Comparing this expression with Eq. (\ref{F3+1}), we can establish a relationship between the 1-forms $\underline{p}^a$ and $\underline{r}^a$ and the 1-forms $\underline{v}^a$ and $\underline{\omega}^a$ as well as the 0-forms $\varphi^a$ and $\chi^a$:
\begin{align}
&\underline{p}^a = \ext{}\varphi^a +\ext{}\varphi^a(u) \underline{u} + \ext{}\underline{v}^a (\cdot,u) - \varrho f^a{}_{bc} (\chi^b \underline{v}^c - \varphi^c \underline{\omega}^b) \,, \label{p3+1} \\
&\underline{r}^a = [\star(\ext{}\underline{v}^a + \varrho f^a{}_{bc} \underline{\omega}^b \wedge \underline{v}^c)] (u,\cdot) \,. \label{r3+1}
\end{align}
The previous two expressions give us relevant information about the behaviour of $\underline{p}^a$ and $\underline{r}^a$. We will consider the cases where a potential term exists and where it does not.

\subsubsection{$V = 0$}

 When the potential term vanishes, just $\mathcal{F}^a$ must be homogeneous and so must be $\underline{p}^a$ and $\underline{r}^a$.  Therefore, Eqs. (\ref{p3+1}) and (\ref{r3+1}) are preserved. \\

 For $V=0$, the expressions in Eq.~(\ref{matter2-f}) reduce to the pure Maxwell form which is \emph{exactly traceless}: $P^\mathcal{K}=\rho^\mathcal{K}/3$ identically, for any $\underline{p}^a,\underline{r}^a$ and regardless of $\varrho$.\footnote{The charge under some non-Abelian Lie group only affects how $\mathcal{F}^a$ relates to $\underline{A}^a$, not this purely algebraic property of the Maxwell stress tensor.} The shear-free continuity equation, Eq.~(\ref{eq:4.6}), then forces the standard radiation dilution $\rho^\mathcal{K}\propto a^{-4}$. Now, since $\rho^\mathcal{K}=\frac{1}{2}(p^2+r^2)$ is built from the same quadratic combination of $\underline{p}^a,\underline{r}^a$ as $\pi^\mathcal{K}$ (Eq.~(\ref{matter2-f})), $\pi^\mathcal{K}$ must dilute the same way: $\pi^\mathcal{K}\propto a^{-4}$ \footnote{The trace-free part of the positive semi-definite tensor $\underline{p}^a\otimes\underline{p}_a+\underline{r}^a\otimes\underline{r}_a$ is bounded by its trace $2\rho^\mathcal{K}$, whatever the isospin orientations are.} which is incompatible with the $a^{-2}$ decay required by Theorem~\ref{1sttheo}.

 Thus the free $1$-form cannot balance the anisotropic curvature when $V=0$, \emph{whether or not it is charged}. This extends Thorsrud's result for the uncharged case~\cite{Thorsrud:2017rpl} to hold also in the charged one: the tracelessness of $\mathcal{T}^\mathcal{K}$ (Eq.~(\ref{matter2-f})) does not involve $\varrho$.

\subsubsection{$V \neq 0$}

If there exists a potential term in the Lagrangian, the 1-form $\underline{A}^a$ must be homogeneous, and so must be $\varphi^a$ and $\underline{v}^a$;  thus, the spatial gradient of $\varphi^a$ drops out of Eq.~(\ref{p3+1}), and $\underline{p}^a$ and $\underline{r}^a$ become
\begin{align}
&\underline{p}^a = \ext{}\underline{v}^a (\cdot,u) - \varrho f^a{}_{bc} (\chi^b \underline{v}^c - \varphi^c \underline{\omega}^b) \,, \label{p3+1a} \\
&\underline{r}^a = [\star(\ext{}\underline{v}^a + \varrho f^a{}_{bc}\, \underline{\omega}^b \wedge \underline{v}^c)] (u,\cdot) \,. \label{r3+1a}
\end{align}
Note that homogeneity does not remove the term $\ext{}\underline{v}^a$ in Eq.~(\ref{r3+1a}): although the components of $\underline{v}^a$ in the orthonormal frame depend on $t$ only, the frame 1-forms are not closed in a Bianchi model, $\ext{}\underline{e}^i\propto c^i{}_{jk}\,\underline{e}^j\wedge\underline{e}^k$, so that $\ext{}\underline{v}^a$ keeps a spatial part proportional to the structure constants. Eq.~(\ref{r3+1a}) thus coincides with Eq.~(\ref{r3+1}); only in Bianchi type I does it reduce to $\varrho f^a{}_{bc}[\star(\underline{\omega}^b\wedge\underline{v}^c)](u,\cdot)$.

Both $\pi^\mathcal{K}$ and $q^\mathcal{K}$ in Eq.~(\ref{matter2-f}) depend on the fields only through isospin-contracted products ($\underline{p}^a\otimes\underline{p}_a$, $\underline{r}^a\otimes\underline{r}_a$, $\underline{v}^a\otimes\underline{v}_a$, $\varphi^2$, $\varphi^a\underline{v}_a$); therefore, a rotation or locking of isospin directions (for instance generated by $\chi^a$) changes nothing which implies we should allow for vectors rotating in isospin space. Thus, it should be possible, in principle, to interpret the condition 
 $
 \pi^\mathcal{K}={}^3S\,
 $
as a condition for tuning $V$'s evolution since it involves $V^\prime(A^2)$. Moreover, we must require
$q^\mathcal{K}=0$ and the continuity equation which, for any $V(A^2)$, reads
\begin{equation}
\tfrac12\dot\Sigma+2H\Sigma+2\big(V'(A^2)\varphi^2\big)^{\boldsymbol\cdot}+V'(A^2)\big(v^2-\varphi^2\big)^{\boldsymbol\cdot}+2HV'(A^2)\big(3\varphi^2+v^2\big)=0\,, \label{cont1+1}
\end{equation}
where $\Sigma\equiv p^2+r^2$; this is a single scalar ordinary differential equation. To demonstrate that this is not simply a mathematical artefact, we illustrate that these conditions can be met. Let us take $\mathcal{B}=0$ (so that $\mathcal{F}^a=\ext{}\underline{A}^a$), $\varphi^a=0$, Bianchi class A with $c^i{}_{jk}=\epsilon_{jki}\,n_i/a$ diagonal (see Refs. \cite{MacCallum:1973kva,Ellis:2012,Zee:2016fuk}), and three orthonormal isovectors $e_{(i)}\in\mathfrak{g}$ with $\underline{A}^a=\sum_i e_{(i)}^a\,(h_i(t)/a)\,\underline{e}^i$. Then, both $\underline{p}^{(i)}\propto\dot h_i/a$ and $\underline{r}^{(i)}\propto-n_ih_i/a^2$ point along $\underline{e}^i$, so $\underline{p}^a\otimes\underline{p}_a$, $\underline{r}^a\otimes\underline{r}_a$, $\underline{v}^a\otimes\underline{v}_a$ are diagonal and $q^\mathcal{K}=0$ identically. The shear-free condition then gives two equations and Eq.~(\ref{cont1+1}) one for the four functions $h_1,h_2,h_3,V'$, which leaves one function free (checked symbolically for general $n_i$). \\

The final conclusion is that a balancing configuration is not excluded but the potential must be tuned to the background $H(t)$.


\subsection{$p=2$ case}

 Let us start with the following action for an underlying differential 2-form $\underline{W}^a$:
\begin{equation}
    S=\int \left[-\frac{1}{2}\mathcal{F}^a\wedge \star \mathcal{F}_a - \star V(W^2) \right] \,.
\end{equation} \\

 According to the 3+1 decomposition\footnote{For further details regarding this decomposition, refer to Appendix~\ref{App:decomposition}.}, we arrive at 
\begin{align}
    &{\cal F}^a = - \star \underline{b}^a + \phi^a\,\star \underline{u} \,, \quad \text{with} \quad \underline{b}^a (u) =0 \,, \label{F2_3+1} \\
    &\underline{W}^a =\underline{u} \wedge \underline{p}^a+\star (\underline{u} \wedge \underline{r}^a) \,, \quad \text{with} \quad \underline{p}^a (u) =0 \quad\text{and}\quad \underline{r}^a (u) =0 \,. \label{W3+1}
\end{align}
In these expressions, $\underline{b}^a = \{\star [\mathcal{F}^a(u,\cdot,\cdot)]\} (\cdot,u)$, $\phi^a = - (\star \mathcal{F}^a) (u)$, $\underline{p}^a = \underline{W}^a (\cdot,u)$, and $\underline{r}^a = (\star \underline{W}^a) (u,\cdot)$. \\

 Thus, the energy-momentum tensor associated with this gauge field is given by 
\begin{align}
    \mathcal{T}^\mathcal{K} = &\underline{u} \otimes \underline{u} \left[\frac{1}{2}(\phi^2 + b^2) + 4V'(W^2)p^2 + V(W^2)\right] \nonumber \\
    & - 2\phi^a {\rm Sym}(\underline{u} \otimes \underline{b}_a) +8V'(W^2)\ {\rm Sym}\{\underline{u} \otimes [\star(\underline{p}^a \wedge \underline{r}_a)](u,\cdot)\} \nonumber \\
    &+ h \left[\frac{1}{2}(\phi^2 - b^2) + 4V'(W^2)r^2 - V(W^2) \right] \nonumber \\
    &+ \underline{b}^a \otimes \underline{b}_a - 4V'(W^2)\left(\underline{p}^a \otimes \underline{p}_a +\underline{r}^a \otimes \underline{r}_a \right) \,, \label{T3-form}
\end{align}
where $b^2 \equiv \underline{b}^a (b_a)$, $b_a$ being the vector dual to $\underline{b}_a$ by metric $g$-duality, and the prime means a derivative with respect to $W^2 \equiv \underline{W}^a (W_a)$, $W_a$ being the vector dual to $\underline{W}_a$ by metric $g$-duality. \\

 Hence, the energy density, pressure, heat flux, and anisotropic stress
associated with the energy-momentum tensor in Eq. (\ref{T3-form}) are the following: 
\begin{align}
    \rho^\mathcal{K} &=\frac{1}{2}(\phi^2 + b^2) + 4V'(W^2)p^2 + V(W^2)  \,, \nonumber \\
    P^\mathcal{K} &=\frac{1}{2}\left(\phi^2 - \frac{1}{3}b^2\right) + \frac{4}{3}V'(W^2)(-p^2 + 2r^2) - V(W^2) \,, \nonumber \\
    q^\mathcal{K} &=  -\phi^a \underline{b}_a +4 V'(W^2) [\star(\underline{p}^a \wedge \underline{r}_a)](u,\cdot)  \,, \nonumber \\
    \pi^\mathcal{K} &= \underline{b}^a \otimes \underline{b}_a - 4V'(W^2)\left(\underline{p}^a \otimes \underline{p}_a +\underline{r}^a \otimes \underline{r}_a \right) + h \left[-\frac{1}{3} b^2 + \frac{4}{3}V'(W^2)(p^2 + r^2) \right]   \,. \label{matter3-f}
\end{align} \\

 Eq. (\ref{PdK}), in this case, is given by
\begin{equation}
\mathcal{F}^a \equiv \ext{}\underline{W}^a + \varrho f^{a}{}_{bc}\,\mathcal B^b\wedge \underline{W}^c \,, \label{Fcd}
\end{equation}
where the connection on the fibre, $\mathcal{B}$, as we already know, is given by
\begin{equation}
\mathcal{B}^a = - \chi^a \underline{u} + \underline{\omega}^a \,, \quad \text{with} \quad \underline{\omega}^a (u) = 0 \,. \label{B3+1a}
\end{equation}
Combining Eqs. (\ref{W3+1}), (\ref{Fcd}), and (\ref{B3+1a}), we end up with
\begin{align}
\mathcal{F}^a = \underline{u} \wedge \{-\ext{}\underline{p}^a-\varrho f^a{}_{bc} [\chi^b \star (\underline{u} \wedge \underline{r}^c) + \underline{\omega}^b \wedge \underline{p}^c]\} + \ext{} \star (\underline{u} \wedge \underline{r}^a) + \varrho f^a{}_{bc} \underline{\omega}^b \wedge \star (\underline{u} \wedge \underline{r}^c) \,.
\end{align}
Comparing this expression with Eq. (\ref{F2_3+1}), we can establish a relationship between the set 0-form $\phi^a$ and 1-form $\underline{b}^a$ and the set 0-form $\chi^a$ and 1-forms $\underline{p}^a$, $\underline{r}^a$, and $\underline{\omega}^a$:
\begin{align}
&\phi^a = -\{\star [ \ext{}\star(\underline{u} \wedge \underline{r}^a)]\}(u) - \varrho f^a{}_{bc} \underline{\omega}^b(r^c) \,, \label{phi3+1} \\
&\underline{b}^a = (\star\{\ext{}\underline{p}^a +[\ext{}\star(\underline{u} \wedge \underline{r}^a)](u,\cdot,\cdot) + \varrho f^a{}_{bc}  \underline{\omega}^b \wedge \underline{p}^c\})(\cdot,u) - \varrho f^a{}_{bc} \chi^b \underline{r}^c \,, \label{b3+1}
\end{align}
where $r^a$ is the vector dual to $\underline{r}^a$ by metric $g$-duality.
The previous two expressions give us relevant information about the behaviour of $\phi^a$ and $\underline{b}^a$. As for the other $p$-form cases, we will consider the situations where a potential term exists and where it does not.

\subsubsection{$V = 0$}

 When the potential term vanishes, just $\mathcal{F}^a$ must be homogeneous and $\phi^a$ and $\underline{b}^a$ must be too.  Therefore, Eqs. (\ref{phi3+1}) and (\ref{b3+1}) are preserved. \\

 To make $q^\mathcal{K} = 0$, Eq. (\ref{matter3-f}) tells us that $\phi^a \perp \underline{b}_a$ in the isospin space. Regarding $\pi^\mathcal{K}$, its required evolution $\pi^\mathcal{K} \propto a^{-2}$ implies $\underline{b}^a \propto a^{-1}$.

 Let us study the free case in the first place, i.e., let us assume $\varrho = 0$.  Eqs. (\ref{phi3+1}) and (\ref{b3+1}) then reduce to
 \begin{align}
&\phi^a = -\{\star [ \ext{}\star(\underline{u} \wedge \underline{r}^a)]\}(u) \,, \label{phi3+1free} \\
&\underline{b}^a = (\star\{\ext{}\underline{p}^a +[\ext{}\star(\underline{u} \wedge \underline{r}^a)](u,\cdot,\cdot) \})(\cdot,u) \,. \label{b3+1free}
\end{align} \\

As we can see, the terms in the gauge-field continuity equation evolve like
\begin{align}
&\dot{\rho}^\mathcal{K} = -b_0^2 \frac{\dot{a}}{a^3} + \phi^a \dot{\phi}_a  \,, \label{2frho} \\
&3H(\rho^\mathcal{K} + P^\mathcal{K}) = b_0^2 \frac{\dot{a}}{a^3} + 3 \frac{\dot{a}}{a} \phi^2  \,. \label{2fP}
\end{align}
By requiring $\phi^a = 0$,\footnote{For instance, by making $\underline{\dot{p}}^a = \underline{r}^a = 0$ in Eqs. (\ref{phi3+1free}) and (\ref{b3+1free}).} leaving untouched $\underline{b}^a$, the gauge-field continuity equation can be satisfied; moreover, the $\phi^a \perp \underline{b}_a$ condition in the isospin space to make $q^\mathcal{K} = 0$ is satisfied as well.  The other possibility to satisfy the gauge-field continuity equation is connected to a non-trivial solution of the differential equation 
\begin{equation}
\phi^a (\dot{\phi}_a + 3H \phi_a) = 0 \,, \label{2fde}
\end{equation}
under the constraints $\phi^a \neq 0$ and Eqs. (\ref{phi3+1free}) and (\ref{b3+1free}); 
however, this would imply a specific evolution for the expansion parameter that is impossible to satisfy for an arbitrary matter content in the $\Lambda$CDM fluid.   The similarity in the expressions and the same conclusion in comparison with the findings in the free 0-form case remind us of the Hodge duality between these two cases discussed in Ref. \cite{Freedman:2012zz} and well exploited by Throsrud in Ref. \cite{Thorsrud:2017rpl}.   \\

Now, let us investigate the non-free case.  
 The same argument as below Eqs. (\ref{2frho}) and (\ref{2fP}) applies so that we can conclude that the shear-free cosmology is realizable only for the $\phi^a = 0$ case and at the expense of some level of fine tuning because of Eqs. (\ref{phi3+1}) and (\ref{b3+1}).\footnote{For instance, by making $\underline{\dot{p}}^a = \underline{r}^a = 0$ and $\underline{\omega}^a \propto a^{-1}$.}  \\

 We can conclude that for a 2-form, uncharged under any non-Abelian Lie group and with no potential, the shear-free cosmology is realizable with no fine tuning; this result is consistent with Thorsrud's findings in Ref. \cite{Thorsrud:2017rpl} in agreement with the Hodge duality between the 0- and 2-form cases discussed in Ref. \cite{Freedman:2012zz}.    
 In contrast, when there exist self interactions, some level of fine tuning is required to realize the balance although there might be instances where the Hodge duality does not hold; indeed, such a duality is not expected to continue holding when introducing general interactions \cite{Plantier:2025hcm,Plantier:2026cgj}.
 

\subsubsection{$V \neq 0$}

If there exists a potential term in the action, the 2-form $\underline{W}^a$ must be homogeneous, and $\underline{p}^a$ and $\underline{r}^a$ must be too;  thus, the spatial derivatives of their components vanish but, as for $p=1$, the exterior derivatives in Eqs.~(\ref{phi3+1}) and~(\ref{b3+1}) do not, since the frame 1-forms are not closed. These two relations therefore hold unchanged. For homogeneous $\underline{r}^a$, the first term in Eq.~(\ref{phi3+1}) only involves the trace $c^j{}_{ij}$ of the structure constants and therefore vanishes in Bianchi class A, whereas the term $\ext{}\underline{p}^a$ in Eq.~(\ref{b3+1}) keeps a spatial part proportional to the structure constants in any Bianchi model other than type I.

 As for $p=1$, we make no assumption on the form of $V$, on the isospin directions, or on how $\underline{b}^a,\underline{p}^a,\underline{r}^a$ scale separately: $\pi^\mathcal{K}$ and $q^\mathcal{K}$ in Eq.~(\ref{matter3-f}) depend on the fields only through isospin-contracted products, so a rotation or locking of isospin directions changes nothing. Thus, the continuity equation reads\footnote{We have used $W^2=2(r^2-p^2)$ for the norm of the 2-form.} reads 
\begin{align}
\begin{aligned}
\phi^a(\dot\phi_a+3H\phi_a)+\underline{b}^a(\dot{\underline{b}}_a+H\underline{b}_a)+16\,V''(W^2)\,p^2\big[\underline{r}^a(\dot{\underline{r}}_a)-\underline{p}^a(\dot{\underline{p}}_a)\big]& \\
+4V'(W^2)\big[\underline{p}^a(\dot{\underline{p}}_a+2H\underline{p}_a)+\underline{r}^a(\dot{\underline{r}}_a+2H\underline{r}_a)\big]&=0\,. \label{cont2+1}
\end{aligned}
\end{align}
This is a single scalar ordinary differential equation and the expression for $\pi^\mathcal{K}$ in Eq.~(\ref{matter3-f}) fixes $V'(W^2(t))$ along the solution in terms of the fields; $V(W^2)$ then follows.\\

 The relation $\mathcal{F}^a=D\underline{W}^a$, Eqs.~(\rev{\ref{phi3+1}}) and~(\rev{\ref{b3+1}}), will not obstruct this construction, since $\underline{b}^a$ is built from $\underline{r}^a$. The unknown functions $\underline{p}^a$, $\underline{r}^a$ (and $\underline{\omega}^a,\chi^a$) outnumber the conditions (five in the expression for $\pi^\mathcal{K}$ in Eq.~(\ref{matter3-f}), three in the expression for $q^\mathcal{K}$ in the same equation, and one in Eq.~(\ref{cont2+1})), so this should in principle be solvable. One simple instance, in Bianchi class A, is $\underline{\omega}^a=\chi^a=0$ (so $\mathcal{F}^a=\ext{}\underline{W}^a$ for any $\varrho$), $\underline{p}^a=0$, $\phi^a=0$ and three orthonormal isovectors with $\underline{r}^a$ along one spatial direction each. Then $q^\mathcal{K}=0$ holds identically, and the expression for $\pi^\mathcal{K}$ in Eq.~(\ref{matter3-f}) together with Eq.~(\ref{cont2+1}) leave one function free. Hence the potential must be tuned to the background $H(t)$, but a balancing configuration is not excluded. \\

\subsection{$p=3$ case}

 Let us start with the following action for an underlying differential 3-form $\underline{H}^a$:
\begin{equation}
    S=\int \left[-\frac{1}{2}\mathcal{F}^a\wedge \star \mathcal{F}_a - \star V(\underline{H}^2) \right] \,.
\end{equation} \\

 According to the 3+1 decomposition\footnote{For further details regarding this decomposition, refer to Appendix~\ref{App:decomposition}.}, we arrive at 
\begin{align}
    &\mathcal{F}^a =\underline{u} \wedge \psi^a \star \underline{u}  \,, \label{F3_3+1} \\
    &\underline{H}^a =-\star \underline{b}^a + \phi^a\,\star \underline{u} \,, \quad \text{with} \quad \underline{b}^a (u) =0 \,. \label{H3+1}
\end{align}
In these expressions, $\psi^a = \{\star [\mathcal{F}^a(u,\cdot,\cdot,\cdot)]\} (u)$,  $\underline{b}^a = \{\star [\underline{H}^a(u,\cdot,\cdot)]\} (\cdot,u)$, and $\phi^a = -(\star \underline{H}^a)(u)$. 
\\

 Thus, the 3+1 decomposition enables us to write the energy-momentum tensor associated with the gauge 3-form field as 
\begin{align}
    \mathcal{T}^\mathcal{K} = &\underline{u} \otimes \underline{u} \left[\frac{1}{2} \psi^2  + 12 V'(\underline{H}^2) b^2 + V(\underline{H}^2) \right] \nonumber \\
    & -24V'(\underline{H}^2) \phi^a {\rm Sym} (\underline{u} \otimes \underline{b}_a) \nonumber \\
    &+ h \left[-\frac{1}{2} \psi^2  - 12V'(\underline{H}^2) (b^2 - \phi^2) - V(\underline{H}^2) \right] \nonumber \\
    & + 12V'(\underline{H}^2) \underline{b}^a \otimes \underline{b}_a \,, \label{T4-form}
\end{align}
where $\psi^2 \equiv \psi^a \psi_a$, $b^2 \equiv \underline{b}^a (b_a)$, $b_a$ being the vector dual to $\underline{b}_a$ by metric $g$-duality,  $\phi^2 \equiv \phi^a \phi_a$,  and the prime means a derivative with respect to $\underline{H}^2 \equiv \underline{H}^a (H_a)$, $H_a$ being the vector dual to $\underline{H}_a$ by metric $g$-duality. \\

 Hence, the energy density, pressure, heat flux, and anisotropic stress
associated with the energy-momentum tensor in Eq. (\ref{T4-form}) are the following: 
\begin{align}
    \rho^\mathcal{K} =&\frac{1}{2} \psi^2  + 12 V'(\underline{H}^2) b^2 + V(\underline{H}^2)  \,, \nonumber \\
    P^\mathcal{K} =& -\frac{1}{2} \psi^2  - 8V'(\underline{H}^2) \left(b^2 - \frac{3}{2}\phi^2\right) - V(\underline{H}^2)  \,, \nonumber \\
    q^\mathcal{K} =&  -12V'(\underline{H}^2) \phi^a \underline{b}_a  \,, \nonumber \\
    \pi^\mathcal{K} =& 12V'(\underline{H}^2) \underline{b}^a \otimes \underline{b}_a - h\left[ 4V'(\underline{H}^2) b^2 \right] \,. \label{matter4-f}
\end{align} \\

 Eq. (\ref{PdK}), in this case, is given by
\begin{equation}
\mathcal{F}^a \equiv \ext{}\underline{H}^a + \varrho f^{a}{}_{bc}\,\mathcal B^b\wedge \underline{H}^c \,, \label{FH}
\end{equation}
where the connection on the fibre, $\mathcal{B}$, as we already know, is given by
\begin{equation}
\mathcal{B}^a = - \chi^a \underline{u} + \underline{\omega}^a \,, \quad \text{with} \quad \underline{\omega}^a (u) = 0 \,. \label{B3+1b}
\end{equation}
Combining Eqs. (\ref{H3+1}), (\ref{FH}), and (\ref{B3+1b}), we end up with
\begin{align}
\begin{aligned}
\mathcal{F}^a =& -\ext{} \star \underline{b}^a + \{\ext{} \phi^a + \varrho f^a{}_{bc} [\phi^c (\cancel{\underline{\omega}^b} - \chi^b \underline{u})]\} \wedge \star \underline{u} - 3H\phi^a\,\underline{u}\wedge\star\underline{u} \\
&+ \varrho f^a{}_{bc} (\cancel{\chi^b \underline{u}} - \underline{\omega}^b)\wedge \star \underline{b}^c \,.
\end{aligned}
\end{align}
The struck-through terms vanish identically, since $\underline{\omega}^b\wedge\star\underline{u}=\underline{\omega}^b(u)\,\eta=0$ and $\underline{u}\wedge\star\underline{b}^c=\underline{b}^c(u)\,\eta=0$, $\eta$ being the volume form; the term $-3H\phi^a\,\underline{u}\wedge\star\underline{u}$ arises from $\phi^a\,\ext{}\star\underline{u}$, which does not vanish for an expanding congruence.
Comparing this expression with Eq. (\ref{F3_3+1}), we can establish a relationship between the 0-form $\psi^a$, on one hand, and the 0-forms $\phi^a$ and $\chi^a$ and the 1-forms $\underline{b}^a$ and $\underline{\omega}^a$, on the other hand:
\begin{align}
\psi^a =& -\{\star [(\ext{} \star \underline{b}^a)(u,\cdot,\cdot,\cdot)]\} (u) - \ext{} \phi^a (u) - 3H\phi^a + \varrho f^a{}_{bc} [-\chi^b \phi^c + \underline{\omega}^b (b^c)] \,, \label{psi3+1} 
\end{align}
This expression gives us relevant information about the behaviour of $\psi^a$. As for the other $p$-form cases, we will consider the situations where a potential term exists and where it does not.

\subsubsection{$V = 0$}

 When the potential term vanishes, just $\mathcal{F}^a$ must be homogeneous and so must be $\psi^a$.  Therefore, Eq. (\ref{psi3+1}) is preserved. \\

 First of all, as we can see from Eq. (\ref{matter4-f}), both $\pi^\mathcal{K}$ and $q^\mathcal{K}$ vanish identically.
On the other hand, the time evolution of the two terms in the left member of the gauge-field continuity equation turns out to be given by
\begin{align}
&\dot{\rho}^\mathcal{K} = \psi^a \dot{\psi}_a \,, \\
&3H(\rho^\mathcal{K} + P^\mathcal{K}) = 0 \,.
\end{align}
This result makes us conclude that we are facing the trivial case of a cosmological constant (as already anticipated in Ref. \cite{Hawking:1984hk}):  there is neither heat flux nor anisotropic stress and the homogeneity condition implies $\psi^a$ to be homogeneous.  The gauge-field continuity equation demands $\psi^a$ to be constant and,   clearly, the cosmological constant is proportional to $\psi^2$. \\


 In conclusion, the free 3-form case is trivial 
 and, therefore, does not lead to a shear-free cosmology where there are anisotropic stress and anisotropic curvature that, being non vanishing, balance each other.

\subsubsection{$V \neq 0$}

 If there exists a potential term in the Lagrangian, the 3-form $\underline{H}^a$ must be homogeneous, and so must be $\underline{b}^a$ and $\phi^a$;  thus, the gradient of $\phi^a$ drops out, and $\psi^a$ becomes
\begin{align}
\psi^a =& -\{\star [(\ext{} \star \underline{b}^a)(u,\cdot,\cdot,\cdot)]\} (u) - \ext{} \phi^a (u) - 3H\phi^a + \varrho f^a{}_{bc} [-\chi^b \phi^c + \underline{\omega}^b (b^c)] \,, \label{psi3+1a} 
\end{align}
For homogeneous $\underline{b}^a$, the terms involving $\ext{}\star\underline{b}^a$ are proportional to the divergence of $\underline{b}^a$, which reduces to a contraction of $\underline{b}^a$ with the trace $c^j{}_{ij}$ of the structure constants; they therefore vanish in Bianchi class A or, more generally, when $\underline{b}^a$ is orthogonal to this trace.

As for $p=1,2$, we make no assumption on the form of $V$ or on how $\underline{b}^a$ scales: both $\pi^\mathcal{K}$ and $q^\mathcal{K}$ in Eq.~(\ref{matter4-f}) depend on $\underline{b}^a$ only through $\underline{b}^a\otimes\underline{b}_a$ and $\phi^a\underline{b}_a$, so a rotation or locking of isospin directions changes nothing. Thus, the continuity equation reads\footnote{We have used $\underline{H}^2=6(\phi^2-b^2)$ for the norm of the 3-form.}
\begin{equation}
\psi^a\dot\psi_a+144\,V''(\underline{H}^2)\,b^2\big[\phi^a\dot\phi_a-\underline{b}^a(\dot{\underline{b}}_a)\big]+12V'(\underline{H}^2)\big[\underline{b}^a(\dot{\underline{b}}_a+H\underline{b}_a)+\phi^a(\dot\phi_a+3H\phi_a)\big]=0\,. \label{cont3+1a}
\end{equation}
The expression for $\pi^\mathcal{K}$ in Eq.~(\ref{matter4-f}) constrains only the trace-free part of $\underline{b}^a\otimes\underline{b}_a$; its isotropic part $V'(\underline{H}^2)b^2$ is left free, so $V'(\underline{H}^2)b^2\propto a^{-2}$ is not required. Furthermore, $\psi^a$ is not independent: Eq.~(\ref{psi3+1a}) gives it in terms of $\phi^a$, $\chi^a$, $\underline{\omega}^a$ and $\underline{b}^a$, and for $\varrho\neq0$ it can be adjusted through $\underline{\omega}^a$.\\

 A simple instance is $\mathcal{B}=0$, $\phi^a=0$, $\psi^a=0$,\footnote{In Bianchi class A, $\psi^a=0$ follows from the former two through Eq.~(\ref{psi3+1a}).} and three orthonormal isovectors with $\underline{b}^a$ along one spatial direction each. Then $q^\mathcal{K}=0$ identically, and the expression for $\pi^\mathcal{K}$ in Eq.~(\ref{matter4-f}) together with Eq.~(\ref{cont3+1a}) leave one function free. \\
 
 One subcase is $V'(\underline{H}^2)=\text{const}$ and $\underline{b}^a\propto a^{-1}$, for which Eq.~(\ref{cont3+1a}) holds for any $a(t)$ with $V$ being a first-degree polynomial in $\underline{H}^2$.  In such a case, $V(\underline{H}^2) = A + B \underline{H}^2$ with $A$ and $B$ being constants.

 The charged case, $\varrho\neq0$, is not excluded either. With $\phi^a=0$ and in Bianchi class A, Eq.~(\ref{psi3+1a}) gives $\psi^a=\varrho f^a{}_{bc}\,\underline{\omega}^b(b^c)$, which vanishes whenever the connection is chosen such that $f^a{}_{bc}\,\underline{\omega}^b(b^c)=0$; this is the case, for instance, if $\underline{\omega}^a$ is aligned with $\underline{b}^a$ in isospin space for each spatial direction, the total antisymmetry of $f_{abc}$ then making the contraction vanish. With this choice of the connection, the configuration described above, including the subcase $V(\underline{H}^2)=A+B\,\underline{H}^2$, carries over unchanged to $\varrho\neq0$. Being uncharged is therefore sufficient, but not necessary, for this balancing mechanism.

 As a conclusion, the 3-form is capable of producing a shear-free cosmology where its anisotropic stress balances the anisotropic curvature. In general, the potential must be tuned to the background, but a balancing configuration is not excluded. A very special and successful configuration, in Bianchi class A, is that where the 3-form is not charged under any non-Abelian Lie group (or is charged, with the connection chosen as described above) and whose potential term is a first-degree polynomial of $\underline{H}^2$.

\subsection{$p=4$ case}

 Let us start with the following action for an underlying differential 4-form $\underline{N}^a$:
\begin{equation}
    S=\int \left[-\star V(N^2)\right] \,.
\end{equation}
The kinetic term is not present in this expression since $\underline{N}$ is a top form. \\

 According to the 3+1 decomposition\footnote{For further details regarding this decomposition, refer to Appendix~\ref{App:decomposition}.}, we know that 
\begin{equation}
    \underline{N}^a =\underline{u} \wedge \psi^a \star \underline{u} \,.
\end{equation}
In this expression, $\psi^a = \{\star [\underline{N}^a(u,\cdot,\cdot,\cdot)]\} (u)$.  \\

 Thus, the energy-momentum tensor associated with this gauge field is given by 
\begin{align}
    \mathcal{T}^\mathcal{K} = &\underline{u} \otimes \underline{u} [48 V'(N^2) \psi^2 + V(N^2)] \nonumber \\
    & - h [48 V'(N^2) \psi^2 + V(N^2)] \,, \label{T5-form}
\end{align}
where $\psi^2 \equiv \psi^a \psi_a$ and the prime means a derivative with respect to $N^2 \equiv \underline{N}^a (N_a)$, $N_a$ being the vector dual to $\underline{N}_a$ by metric $g$-duality. \\

 Hence, the energy density, pressure, heat flux, and anisotropic stress
associated with the energy-momentum tensor in Eq. (\ref{T5-form}) are the following: 
\begin{align}
    \rho^\mathcal{K} &=48 V'(N^2) \psi^2 + V(N^2) \,, \nonumber \\
    P^\mathcal{K} &= -48 V'(N^2)\psi^2 - V(N^2) \,, \nonumber \\
    q^\mathcal{K} &=  0  \,, \nonumber \\
    \pi^\mathcal{K} &= 0   \,. \label{mprop5}
\end{align} \\


 Since $\pi^\mathcal{K}=0$, this is not a candidate for balancing anisotropies. Clearly, because $P^\mathcal{K} = -\rho^\mathcal{K}$ in this case, we are dealing with a cosmological constant, as expected. Indeed, the gauge-field continuity equation reduces to
$\dot{\rho}^\mathcal{K}=0$. \\

\section{Conclusions}
\label{sec:conclusion}

 Astronomical observations at cosmological scales reveal a universe that is highly homogeneous and isotropic and that, therefore, can be modelled through the Friedmann-Lemaitre-Robertson-Walker spacetime \cite{Peter:2013avv,Weinberg:2008zzc,Mukhanov:2005sc,Lyth:2009imm,Ellis:2012}.  Of course, there is always room for small departures from this cosmological principle, this being the starting point for many works that involve inhomogeneous but, specially, anisotropic spacetimes (see Ref. \cite{MacCallum:1973kva} and the respective references in Refs. \cite{Peter:2013avv,Ellis:2012}).  Our concern in this work has not been that of implementing some version of an anisotropic universe but, instead, we have questioned the cornerstone status of the cosmological principle.  Based on the fact that astronomical observations at the background level, (i.e., not considering the particularities of structure formation) just point towards a homogeneous, irrotational, and shear-free spacetime, we have followed previous ideas about implementing homogeneous, irrotational, and shear-free cosmologies that, still, are anisotropic in the spatial curvature.\footnote{This is in contrast to other works, like Ref. \cite{BeltranJimenez:2026bif}, where the cosmological principle is tried to be preserved at all cost.}  \\

 Among the several works that have pursued these ideas, the one by Thorsrud in Ref. \cite{Thorsrud:2017rpl} is, probably, the most recent and most complete.  Thorsrud tries implementing the homogeneous, irrotational, and shear-free cosmologies by making use of a $p$-form that joins the $\Lambda$CDM fluid but whose work is to balance the anisotropic spatial curvature with its anisotropic stress.  This way, a class of eFLRW cosmologies emerges that is as suitable to describe the background evolution of the universe as the FLRW cosmology is \cite{Thorsrud:2019csi}.  Thorsrud's work deals with a free $p$-form that is uncharged under any internal symmetry group (or charged under an Abelian one).  Although Thorsrud demonstrates the 0- and 2-form cases to be successful at doing the job of balancing the anisotropic spatial curvature, they being equivalent via Hodge duality \cite{Freedman:2012zz}, the absence of interactions is a significant drawback that may easily lead to the conclusion that the scenario is highly fine tuned and, therefore, unworthy of further investigation.  \\

 We have tried to remove this drawback by analyzing the effect of introducing self interactions through both charging the $p$-form under a non-Abelian Lie group and the explicit introduction of a potential term for the $p$-form in the action.  The obtained results have allowed us to establish, beyond any doubt, up to what extent the shear-free scenario is fine tuned. \\

 We have checked that, as our expectations demanded, both the existence of a charge under some non-Abelian Lie group and the introduction of a potential term in the action are crucial for the success or failure of the scenario.  However, there exists a third element that turned out to be relevant:  the intimate relation between the gauge-field strength $\mathcal{F}^a$ and the $p$-form $\mathcal{K}^a$.  Such a relation, expressed through the identity $\mathcal{F}^a \equiv D \mathcal{K}^a$, where $D$ is the gauge-covariant exterior derivative, establishes a set of strong relationships between the degrees of freedom that describe both $\mathcal{F}^a$ and $\mathcal{K}^a$. 

 The findings of our work, as clearly summarized in Table \ref{tab:summary},  suggest that, among the alternatives, the $3$-form gauge field is the most interesting candidate. While it is true that also matter sectors built from $p$-forms of other ranks may work, they are flinged with fine-tuning issues. Indeed, the potential must be tuned against the expansion history $a(t)$. The only candidate we have found that does not require a finely tuned form of the potential (against the expansion history) is indeed a $3$-form gauge field with a potential quadratic in the $3$-form: $V(\underline{H}^2)=A+B\,\underline{H}^2$ where $A$ and $B$ are constants; in Bianchi class A, this holds for an uncharged 3-form and, with a suitably chosen connection, also for a charged one. 
 The $3$-form gauge field is thus a natural candidate for dark energy in combination with a dark-matter term mimicking curvature, and it would be interesting to investigate whether it can be used to construct a viable model of the cosmic expansion dark history. Although being able to mimic the FLRW cosmology at the background level, this scenario will show some departures at the perturbative level that are worthy of investigating and contrasting with the observations.  Moreover, the 3-form being an additional matter field that may eventually decay,\footnote{If we suitably couple it to the $\Lambda$CDM matter fluid so that the findings in this paper are not spoiled.} it can inject energy into the $\Lambda$CDM fluid before recombination and constitute an interesting contribution towards the solution to the, well known nowadays, Hubble tension \cite{Wang:2026jxd,Schoneberg:2026eys,Schoneberg:2026vaf,Jia:2026vdt}. These ideas will be the subject of forthcoming work.  

In this work, we have not made any specific assumptions about the Lie group except that it is compact and semisimple. One may thus wonder what the results would look like for particular groups and when faced with the restrictions given by Eqs.~(\ref{eq:FieldEq}), (\ref{eq:IntCond}) and (\ref{eq:J}) applied to particular Bianchi models. This deserves a thorough investigation on its own and is left for future work. In particular, one must investigate the stability of shear-free solutions against anisotropic perturbations.


\ack{Nicolás Hernández gratefully acknowledges Universidad Industrial de Santander for supporting this research through a Master's research scholarship; he also wishes to express his sincere gratitude to Ben David Normann and Yeinzon Rodríguez for their invaluable guidance, dedication, and patience throughout the development of this work. Yeinzon Rodríguez has been funded by Universidad Antonio Nariño under grant numbers VCTI 2024211, VCTI 2025205, VCTI 2232026, and VCTI 2252026.   Ben David Normann and Yeinzon Rodríguez acknowledge John Bayron Orjuela Quintana for his comments and help at early stages of this work.}

\appendix

\section{$3+1$ $p$-form gauge-field decomposition}
\label{App:decomposition}

\begin{thm}[1-form]
    Let ${\cal P}$ be a 1-form. Given a unit time-like vector $u$ (in practice, it will be the 4-velocity of some observer), there exist a unique 0-form $\varphi$ and a unique 1-form $\underline{v}$ such that
\begin{align}
    {\cal P}=-\varphi\,\underline{u}+\underline{v} \,, \quad \text{with} \quad \underline{v} (u) = 0 \,,
\end{align}
where $\underline{u}$ is the 1-form dual to $u$ by metric $g$-duality.
\end{thm}

\subsection*{{\bf Proof}}

It is obvious. Of course, $\varphi = \mathcal{P}(u)$ and $\underline{v} = \mathcal{P} + \mathcal{P}(u) \ \underline{u}$. $\square$

\begin{thm}[2-form]
\it Let ${\cal P}$ be a 2-form.
Given a unit time-like vector $u$ (in practice, it will be the 4 -velocity of some observer), there exist two unique 1-forms $\underline{p}$ and $\underline{r}$ such that
\begin{align}
    {\cal P}=\underline{u} \wedge \underline{p}+\star (\underline{u} \wedge \underline{r}) \,, \quad \text{with} \quad \underline{p} (u) =0 \quad\text{and}\quad \underline{r} (u) =0 \,, \label{2Fdecomp}
\end{align}
where $\underline{u}$ is the 1-form dual to $u$ by metric $g$-duality.
\end{thm}

\subsection*{{\bf Proof}\footnote{The statement and proof of this theorem is found in Ref. \cite{Gourgoulhon:2013gua}.}}

Let us assume the validity of Eq. (\ref{2Fdecomp}).  Introducing $u$ in the second slot of $\mathcal{P}$, we obtain
\begin{align}
\mathcal{P} (\cdot,u) = \underline{u} \ \cancelto{\text{\tiny 0}}{\underline{p}(u)} - \underline{p} \ \underline{u}(u) + \cancelto{\text{\tiny 0}}{\eta(u,r,\cdot,u)} = \underline{p} \,,
\end{align}
where $r$ is the vector dual to the 1-form $\underline{r}$ by metric $g$-duality and $\eta$ is the canonical volume form.
This proves the existence and uniqueness of $\underline{p}$.  Let us now introduce $u$ in the first slot of $\star \mathcal{P}$:
\begin{align}
(\star \mathcal{P}) (u,\cdot) = \cancelto{\text{\tiny 0}}{\eta(u,p,u,\cdot)} - \underline{u}(u) \ \underline{r} + \cancelto{\text{\tiny 0}}{\underline{r}(u)} \ \underline{u}  = \underline{r} \,,
\end{align}
where $p$ is the vector dual to the 1-form $\underline{p}$ by metric $g$-duality.  This proves the existence and uniqueness of $\underline{r}$. $\square$


\begin{thm}[3-form]
\it Let ${\cal P}$ be a 3-form.
Given a unit time-like vector $u$ (in practice, it will be the 4-velocity of some observer), there exist a unique 0-form $\phi$ and a unique 1-form $\underline{b}$ such that
\begin{align}
    {\cal P}= -\star \underline{b} + \phi\,\star \underline{u} \,, \quad \text{with} \quad \underline{b} (u) =0 \,, \label{3Fdecomp} 
\end{align}
where $\underline{u}$ is the 1-form dual to $u$ by metric $g$-duality.
\end{thm}

\subsection*{{\bf Proof}}

Let us assume the validity of Eq. (\ref{3Fdecomp}).  Introducing $u$ in the first slot of $\mathcal{P}$, we obtain
\begin{align}
\mathcal{P}(u,\cdot,\cdot) &= -\eta(u,b,\cdot,\cdot) + \phi \ \cancelto{\text{\tiny 0}}{\eta(u,u,\cdot,\cdot)} \nonumber \\
&= - \star (\underline{u} \wedge \underline{b}) \,.
\end{align}
Thus, $\underline{u} \wedge \underline{b} = \star [\mathcal{P}(u,\cdot,\cdot)]$ and, therefore, introducing $u$ in the second slot of this expression, we obtain
\begin{align}
\{\star [\mathcal{P}(u,\cdot,\cdot)] \}(\cdot,u) = \underline{u} \ \cancelto{\text{\tiny 0}}{\underline{b}(u)} - \underline{b} \ \underline{u}(u) = \underline{b} \,.
\end{align}
This proves the existence and uniqueness of $\underline{b}$.  Regarding $\phi$, it is easy to see that $\star \mathcal{P} =  - \underline{b} + \phi \ \underline{u}$ which, applied on $u$ gives
\begin{align}
-(\star \mathcal{P})(u) - \cancelto{\text{\tiny 0}}{\underline{b}(u)} = - \phi \ \underline{u}(u) = \phi \,.
\end{align}
This, in turn, proves the existence and uniqueness of $\phi$. $\square$    


\begin{thm}[4-form]
\it Let ${\cal P}$ be a 4-form.
If the manifold is four dimensional, and given a unit time-like vector $u$ (in practice, it will be the 4-velocity of some observer), there exists a unique 0-form $\psi$ such that
\begin{align}
    {\cal P}= \underline{u} \wedge \psi \star \underline{u} \,, \label{4Fdecomp} 
\end{align}
where $\underline{u}$ is the 1-form dual to $u$ by metric $g$-duality. Equivalently, if one posits the more general ansatz ${\cal P}= \underline{u} \wedge \psi \star \underline{u} + \underline{\mathcal{M}}$, with $\underline{\mathcal{M}} (u,\cdot,\cdot,\cdot) = 0$, similarly to the 1-, 2-, and 3-form cases above, then necessarily $\underline{\mathcal{M}} \equiv 0$.
\end{thm}

\subsection*{{\bf Proof}}

Since spacetime is four dimensional, the vector space of 4-forms is one dimensional: every 4-form is proportional to the canonical volume form $\eta$. Consequently, if a 4-form $\underline{\mathcal{M}}$ satisfies $\underline{\mathcal{M}} (u,\cdot,\cdot,\cdot) = 0$ for the fixed, non-vanishing vector $u$, it must vanish identically. Indeed, writing $\underline{\mathcal{M}} = \lambda \, \eta$ for some scalar $\lambda$ and choosing an orthonormal tetrad $\{u,e_1,e_2,e_3\}$ adapted to $u$, the multilinearity of $\underline{\mathcal{M}}$ gives $\underline{\mathcal{M}} (\alpha u,e_1,e_2,e_3) = \alpha\, \underline{\mathcal{M}} (u,e_1,e_2,e_3) = 0$ for every $\alpha \in \mathbb{R}$, while $\eta(u,e_1,e_2,e_3) = \pm 1 \neq 0$; since $\underline{\mathcal{M}}(u,e_1,e_2,e_3) = \lambda\,\eta(u,e_1,e_2,e_3)$, this forces $\lambda = 0$, i.e. $\underline{\mathcal{M}} \equiv 0$. This establishes the decomposition in Eq.~(\ref{4Fdecomp}). \\

To determine $\psi$, we introduce $u$ in the first slot of $\mathcal{P}$:
\begin{align}
\mathcal{P}(u,\cdot,\cdot,\cdot) =& \underline{u} (u) \ \psi \star \underline{u} - \psi \ \cancelto{\text{\tiny 0}}{\eta(u,u,\cdot,\cdot)} \wedge \underline{u} + {\rm Trans} [\underline{u} \wedge \psi \ \cancelto{\text{\tiny 0}}{\eta(u,u,\cdot,\cdot)}] \nonumber \\
&- \underline{u} \wedge \psi \ \cancelto{\text{\tiny 0}}{\eta(u,u,\cdot,\cdot)} \nonumber \\
=& -\psi \star \underline{u} \,,
\end{align}
where Trans means transposition in the first two slots. 
Applying $\star$ to both sides, 
we obtain
\begin{equation}
\star[\mathcal{P}(u,\cdot,\cdot,\cdot)] 
= -\psi\,\underline{u} \,.
\end{equation}
Evaluating on $u$ then gives
\begin{equation}
\{\star [\mathcal{P}(u,\cdot,\cdot,\cdot)]\} (u) =  -\psi \ \underline{u}(u) = \psi \,.
\end{equation}
This proves the existence and uniqueness of $\psi$. $\square$

\section{Temporal gauge for the fibre connection}
\label{App:TemporalGauge}

 As explained in Section~\ref{Sec:Nonabelian}, the fibre connection $\mathcal{B}^a$ transforms under a gauge transformation of the bundle in a way that differs from every other adjoint-valued field appearing in this work. Given an infinitesimal gauge parameter $\lambda=\lambda^a\boldsymbol{t}_a$ (a Lie-algebra-valued $0$-form), the connection transforms inhomogeneously:
\begin{equation}
    \delta \mathcal{B}^a=-\frac{1}{\varrho}\ext{}\lambda^a-f^a{}_{bc}\,\mathcal{B}^b\lambda^c \,, \label{eq:gaugeB}
\end{equation}
this being the standard non-Abelian gauge transformation compatible with the covariant derivative of Eq.~(\ref{eq:DX}) (see, e.g., Ref. \cite{Weinberg:1996kr}). \\

 Given the $3+1$ decomposition of $\mathcal{B}^a$ relative to $u$, Eq.~(\ref{B3+1}),
\begin{equation}
    \mathcal{B}^a=-\chi^a\,\underline{u}+\underline{\omega}^a \,, \quad \text{with} \quad \underline{\omega}^a(u)=0 \,,
\end{equation}
we show below that the $0$-form $\chi^a$ carries no gauge-invariant content: it can always be set to zero by an appropriate choice of gauge, regardless of the gauge group or its structure constants. This is the non-Abelian generalization of the familiar ``temporal gauge'' (or Weyl gauge) of ordinary Yang-Mills theory.

\begin{thm}[Temporal gauge]
Let $\mathcal{B}$ be the fibre connection, decomposed as above. Then, for any smooth $\chi^a(t,\vec{x})$, there exists a gauge transformation under which $\chi^a=0$ identically. \label{thm:tempgauge}
\end{thm}

\subsection*{{\bf Proof}}

 Projecting Eq.~(\ref{eq:gaugeB}) along $u$, and using $\chi^a=\mathcal{B}^a(u)$ together with $\ext{}\lambda^a(u)\equiv -\dot\lambda^a$, we obtain the transformation law of $\chi^a$ itself:
\begin{equation}
    \delta \chi^a=\frac{1}{\varrho}\dot\lambda^a - f^a{}_{bc}\chi^b\lambda^c \,.
\end{equation}
We look for a gauge parameter $\lambda^a(t,\vec{x})$ such that the transformed $\chi'^a\equiv\chi^a+\delta\chi^a$ vanishes identically. This amounts to solving, along each integral curve of $u$ (i.e., along each fundamental observer's worldline, parametrized by $t$, at fixed $\vec{x}$),
\begin{equation}
    \dot\lambda^a(t,\vec{x}) = \varrho\left[-\chi^a(t,\vec{x}) + f^a{}_{bc}\,\chi^b(t,\vec{x})\,\lambda^c(t,\vec{x})\right] \,. \label{eq:odegauge}
\end{equation}
For fixed $\vec{x}$, Eq.~(\ref{eq:odegauge}) is a first-order linear ordinary differential equation for $\lambda^a(t)$, with smooth, bounded coefficients on any finite time interval (given smooth $\chi^a(t,\vec{x})$). By the standard existence-and-uniqueness theorem for linear ordinary differential equations (see, e.g., Ref. \cite{lee2003introduction}), it admits a unique solution for arbitrary initial data $\lambda^a(t_0,\vec{x})$ prescribed on a reference slice $\Sigma_{t_0}$ (e.g., $\lambda^a(t_0,\vec{x})=0$). Since this construction holds at every $\vec{x}$, and no property of the structure constants $f^a{}_{bc}$ or of the Lie group metric $G_{ab}$ was used beyond their smoothness, it goes through for any gauge group and any $\chi^a(t,\vec{x})$. Extending $\lambda$ to a finite gauge transformation by the usual path-ordered exponential along $u$ completes the proof. $\square$

\subsection*{{\bf Remark}}

 The residual gauge freedom that preserves $\chi^a=0$ is exactly the subgroup of transformations with $\dot\lambda^a=0$, i.e., time-independent (purely spatial) transformations, under which $\underline\omega^a$ transforms as
\begin{equation}
    \delta\underline\omega^a=-\frac{1}{\varrho}\ext{}\lambda^a-f^a{}_{bc}\,\underline\omega^b\lambda^c \,.
\end{equation}
Hence $\underline\omega^a$ retains exactly the freedom of an ordinary (purely spatial) Yang-Mills gauge field on each slice $\Sigma_t$, and can be set to zero throughout spacetime only if the associated spatial curvature $\ext{}\underline\omega^a+\frac{\varrho}{2}f^a{}_{bc}\,\underline\omega^b\wedge\underline\omega^c$ vanishes identically, i.e., only if the connection carries no genuine spatial (``magnetic'') content. Barring that special case, $\underline\omega^a$ is physical and cannot be removed by a further gauge choice. \\

 In summary: $\chi^a$ is pure gauge and can always be set to zero without loss of generality; $\underline\omega^a$ cannot, in general. 

\section{Proof of Theorem \ref{1sttheo}} \label{App:Knownth}

{\bf Statement:}
(Anisotropic spatial curvature decay in Bianchi models with conformal
expansion.) {\it The components of the anisotropic spatial curvature ${}^{3}S$
relative to an orthonormal basis exhibit a decay proportional to $1/a^2$ in
the shear-free limit, where $a(t)$ denotes the scale factor governing the
conformal expansion.}

\subsection*{{\bf Proof}}
The statement and proof of this theorem are given in the arXiv version 2 of Ref.
\cite{Thorsrud:2017rpl}. Written in terms of components, the line element for a Bianchi-type
cosmological model undergoing conformal expansion is given by \cite{Zee:2016fuk}
\begin{equation}
    ds^2=-dt^2+a^2(t)\,\tilde{h}_{ab}(x^c)\,dx^a dx^b \,, \label{eq:A-metric}
\end{equation}
where $\tilde{h}(x^c)$ is a function of the spatial coordinates $x^c$
alone.\footnote{The Latin indices $a,b,c$ are used in this appendix to denote
spatial coordinates. They are not internal group indices in contrast to the
main part of the manuscript.}
Then, it follows that ${}^{3}\tilde{\rm R}{\rm icci}$ also depends
exclusively on these spatial coordinates. The induced metric on the
3D slices, $h$, is clearly related to $\tilde{h}$ via a conformal
transformation:\footnote{The Greek indices are used here, as in the main part
of the manuscript, as space-time indices.}
\begin{equation}
    h(x^\mu)=a^2(t)\tilde{h}(x^a) \,.
    \label{A.2}
\end{equation}

\noindent In the following, we shall use hats to denote the components of
${}^{3}\mathrm{Ricci}$ and ${}^{3}\tilde{\rm R}{\rm icci}$ relative to an
orthonormal basis, i.e.,
${}^{3}\mathrm{Ricci}_{\hat a\hat b}$ and
${}^{3}\tilde{\rm R}{\rm icci}_{\hat a\hat b}$.
Let $\Lambda^a{}_{\hat a}(x^\mu)$ and
$\tilde{\Lambda}^a{}_{\hat a}(x^c)$ be the components of the respective
basis transformation matrices. Thus, by definition
\begin{align}
    \tilde{\Lambda}^{a}{}_{\hat a}\,
    \tilde{h}_{ab}\,
    \tilde{\Lambda}^{b}{}_{\hat b}
    &=\delta_{\hat a\hat b} \,, \\
    \Lambda^{a}{}_{\hat a}\,
    h_{ab}\,
    \Lambda^{b}{}_{\hat b}
    &=\delta_{\hat a\hat b} \,.
\end{align}

\noindent From the two previous expressions and Eq.~(\ref{A.2}), it is clear that
\begin{equation}
    \Lambda^{a}{}_{\hat a}(x^\mu)
    =\frac{1}{a(t)}
    \tilde{\Lambda}^{a}{}_{\hat a}(x^c) \,. \label{eq:vielbein-scaling}
\end{equation}
In addition, since $\partial_c a(t)=0$, ${}^{3}\mathrm{Ricci}$ remains
invariant under this conformal transformation, i.e.,
${}^{3}\mathrm{Ricci}={}^{3}\tilde{\rm R}{\mathrm{icci}}$.
Therefore, the components ${}^{3}\mathrm{Ricci}_{ab}$ and
${}^{3}\tilde{\rm R}{\mathrm{icci}}_{ab}$ are equal and time independent.
Thus, it follows that
\begin{equation}
    {}^{3}\mathrm{Ricci}_{\hat a\hat b}(x^\mu)
    =
    \frac{{}^{3}\tilde{\rm R}{\rm icci}_{\hat a\hat b}(x^c)}{a^2(t)} \,,
\end{equation}
where the numerator on the right member of the equation is manifestly time independent, being built entirely out of $\tilde h_{ab}(x^c)$ and $\tilde{\Lambda}^a{}_{\hat a}(x^c)$.
This establishes the $1/a^2(t)$ decay of the components of
${}^{3}\mathrm{Ricci}$ relative to an orthonormal basis. Consequently, the
same behaviour holds for the trace-free part ${}^{3}S$. $\square$

\section{Orthonormal tetrad adapted to $u$}
\label{App:Tetrad}

 Throughout this work, all frame indices (those on space-time objects, as opposed to the internal gauge-group indices carried by $\alpha^a$-type fields) are components relative to a non-coordinate, orthonormal tetrad adapted to the fundamental-observer congruence $u$. This appendix makes that tetrad explicit, building directly on the vielbein relation of Appendix~\ref{App:Knownth}, so that the scaling factors invoked in Sec.~\ref{sec:non-Abelian} (and elsewhere) can be read off directly rather than re-derived each time. \\

 For a Bianchi-type metric undergoing conformal expansion, Eq.~(\ref{eq:A-metric}),
\begin{equation}
    ds^2=-dt^2+a^2(t)\,\tilde{h}_{ab}(x^c)\,dx^a dx^b \,,
\end{equation}
the timelike leg of the tetrad is simply the coordinate vector field
\begin{equation}
    u=\partial_t \,, \qquad \underline{u}=-dt \,,
\end{equation}
the sign in $\underline u$ following from the mostly-plus signature, $g(u,u)=g_{tt}=-1$. Since the metric has no $dt\,dx^a$ cross-term, $g(u,\partial_a)=0$ for every spatial coordinate direction, so $u$ is automatically orthogonal to the spatial slices $\Sigma_t$. \\

 The three spatial legs are built from the vielbein of Appendix~\ref{App:Knownth}. Recall that $\tilde{\Lambda}^a{}_{\hat a}(x^c)$ is defined by $\tilde{\Lambda}^{a}{}_{\hat a}\tilde{h}_{ab}\tilde{\Lambda}^{b}{}_{\hat b}=\delta_{\hat a\hat b}$, i.e., it is the (purely spatial-coordinate-dependent, manifestly time-independent) vielbein of the time-independent metric $\tilde h_{ab}(x^c)$. Let us define the vector fields
\begin{equation}
    e_{\hat a} \equiv \Lambda^{a}{}_{\hat a}(x^\mu)\,\partial_a = \frac{1}{a(t)}\,\tilde{\Lambda}^{a}{}_{\hat a}(x^c)\,\partial_a \,, \label{eq:tetrad-vectors}
\end{equation}
using the vielbein-scaling relation in Eq.~(\ref{eq:vielbein-scaling}). By construction, $g(e_{\hat a},e_{\hat b})=\Lambda^{a}{}_{\hat a}\Lambda^{b}{}_{\hat b}h_{ab}=\delta_{\hat a\hat b}$ and $g(u,e_{\hat a})=0$, so $\{u,e_{\hat 1},e_{\hat 2},e_{\hat 3}\}$ is precisely the orthonormal tetrad adapted to $u$ used throughout this work (cf. the tetrad $\{u,e_1,e_2,e_3\}$ used in Appendix~\ref{App:decomposition}). The associated dual coframe is
\begin{equation}
    \hat{e}^{\hat a} = a(t)\,\tilde{\Lambda}^{\hat a}{}_{a}(x^c)\,dx^a \,, \label{eq:tetrad-coframe}
\end{equation}
where $\tilde{\Lambda}^{\hat a}{}_{a}$ is the matrix inverse of $\tilde{\Lambda}^{a}{}_{\hat a}$, i.e., $\tilde{\Lambda}^{\hat a}{}_{a}\tilde{\Lambda}^{a}{}_{\hat b}=\delta^{\hat a}_{\hat b}$; the overall factor of $a(t)$ in Eq.~(\ref{eq:tetrad-coframe}), rather than $1/a(t)$ as in Eq.~(\ref{eq:tetrad-vectors}), is exactly what is required for $\hat{e}^{\hat a}(e_{\hat b})=\delta^{\hat a}_{\hat b}$ to hold. \\

 Eq.~(\ref{eq:tetrad-vectors}) gives the general formula for the orthonormal-frame directional derivative (gradient component) of a scalar $\phi(t,\vec{x})$:
\begin{equation}
    (\ext{}\phi)_{\hat a} \equiv e_{\hat a}(\phi) = \frac{1}{a(t)}\,\tilde{\Lambda}^{b}{}_{\hat a}(x^c)\,\partial_b\phi \,. \label{eq:grad-formula}
\end{equation}
Eq.~(\ref{eq:grad-formula}) is the key formula used in Section~\ref{sec:non-Abelian}: it shows explicitly that {\it any} scalar's orthonormal-frame spatial gradient carries one power of $1/a(t)$ coming purely from the tetrad itself, on top of whatever explicit time-dependence is already present in $\partial_b\phi$. \\

 For concreteness, and to make Eq.~(\ref{eq:grad-formula}) immediately usable, let us specialize to the flat (Bianchi~I) case, $\tilde{h}_{ab}=\delta_{ab}$ in Cartesian coordinates $x^a=(x^1,x^2,x^3)$, for which $\tilde{\Lambda}^a{}_{\hat a}=\delta^a_{\hat a}$ is simply the identity. The tetrad of Eqs.~(\ref{eq:tetrad-vectors})-(\ref{eq:tetrad-coframe}) then reduces to
\begin{equation}
    u=\partial_t \,, \quad e_{\hat\imath}=\frac{1}{a(t)}\,\partial_{x^i} \,, \qquad \underline{u}=-dt \,, \quad \hat{e}^{\hat\imath}=a(t)\,dx^i \,, \label{eq:tetrad-flat}
\end{equation}
for $i=1,2,3$, and Eq.~(\ref{eq:grad-formula}) becomes simply
\begin{equation}
    (\ext{}\phi)_{\hat\imath} = \frac{1}{a(t)}\,\partial_{x^i}\phi \,. \label{eq:grad-formula-flat}
\end{equation}
This is the explicit form of the gradient used, e.g., in Sec.~\ref{sec:non-Abelian} when checking the scaling of $\underline{v}^a$ under different ans\"atze for the $0$-form $\phi^a$: any envelope $\phi^a=\phi^a(t)/a(t)$ that is itself already scaled by $1/a(t)$ picks up a second such factor from Eq.~(\ref{eq:grad-formula-flat}) and gives $\nabla\phi^a\propto a^{-2}$, whereas a spatial profile with no explicit time-dependence gives $\nabla\phi^a\propto a^{-1}$ directly, with no further tuning required. \\

 Finally, note for completeness that the time leg reproduces the familiar time derivative, $u(\phi)=\dot\phi$, and that $\dot{e}_{\hat a}=-H\,e_{\hat a}$ along $u$ for any fixed spatial coordinate (immediate from differentiating Eq.~(\ref{eq:tetrad-vectors}) at fixed $x^a$, since $\tilde\Lambda^a{}_{\hat a}(x^c)$ carries no time dependence) -- the tetrad-frame statement underlying the relations $\dot{\underline v}^a=-H\underline v^a$ used repeatedly in Sec.~\ref{sec:non-Abelian}.


\bibliographystyle{JHEP}
\bibliography{refs}

\end{document}